\documentclass[aps,pra,reprint,unsortedaddress,amsmath,amssymb,floatfix]{revtex4-2}

\usepackage{lipsum}
\usepackage{comment}
\usepackage{xcolor}

\usepackage{graphicx}
\usepackage{dcolumn}

\usepackage{braket}
\usepackage{booktabs}
\usepackage{bm}

\begin{document}


\title{Complex scaling calculation of phase shifts for positron collisions with positive ions}


\author{Taishi Sano}
\altaffiliation[Present address: ]{Department of Physics, Waseda University, Shinjuku 169-8050, Japan}
\affiliation{Department of Chemistry, Tohoku University, Sendai, 980-8578, Japan}

\author{Takuma Yamashita}
\email{tyamashita@tohoku.ac.jp}
\affiliation{Institute for Excellence in Higher Education, Tohoku University, Sendai 980-8576, Japan}
\affiliation{Department of Chemistry, Tohoku University, Sendai, 980-8578, Japan}

\author{Yasushi Kino}%
\affiliation{Department of Chemistry, Tohoku University, Sendai, 980-8578, Japan}


\date{\today}

\begin{abstract}
We present phase-shift calculations for positron collisions with positive ions using a complex scaling method (CSM).
Based on the findings of this study [R. Suzuki, T. Myo, and K. Kat\={o}, Prog. Theor. Phys. \textbf{113}, 1273 (2005).],
we propose a modification of the phase shift in the CSM calculation, in which 
phase shifts are derived only from the complex eigenenergies of the CSM Hamiltonian.
This modification is based on the fact that the contributions of high-lying complex eigenenergies can be approximated 
as a constant value in the case of a small collision energy,
where neither target excitation nor positronium formation occurs.
The proposed modification limits the contribution of the complex eigenenergies to the vicinity of the collision energy,
which is also intuitively acceptable.
We present a geometrical formulation of the modification and demonstrative calculations of positron scattering off positive ions.
Our results agree well with those reported in the literature for the targets Ne, Ar, Kr, Xe, H, He, He$^+$, and Li$^{2+}$.
The phase shifts of positron scattering off a Li$^+$ ion have also been reported.
\end{abstract}


\maketitle

\section{Introduction}

An observation of a positron scattering off an atom provides essential insights into the behavior of positrons in matter. Positrons interact attractively with electrons comprising the target atom and can form a positronium (Ps) atom, which is a bound state of the electron and positron. 
Even when the collision energy is below the threshold for Ps formation, it is crucial to consider virtual Ps formation during the scattering process, which results in a challenging two-center problem and requires rigorous theoretical treatment~\cite{PhysRevA.66.012710,Fursa_2012}.
In recent years, the improvement in the positron beam quality~\cite{doi:10.1063/1.4913354,doi:10.1063/1.4939854} 
has revealed scattering cross-sections with a variety of atoms/molecules 
over a wide range of collision energies~\cite{RevModPhys.82.2557,Chiari:2014vs,Ratnavelu2019}.
Furthermore, the development of Ps beams~\cite{10.1063/1.5128012,PhysRevA.100.013410,doi:10.1063/1.5060619,Nagashima_2021} has stimulated interest in positronium-atom interactions ~\cite{Brawley789,PhysRevLett.115.223201,PhysRevA.90.052717,PhysRevA.97.012706,PhysRevA.105.052812,Yamashita:2021ue,Yamashita_2021}.

Compared to positron scattering off neutral atoms, the investigation of the scattering off positive ions has been relatively limited. 
This is due to experimental difficulties and theoretical complexities arising from the long-range Coulomb repulsion between the positron and ion. 
For instance, positron scattering off a He$^{+}$ ion has been studied using various theoretical approaches~\cite{doi:10.1143/JPSJ.31.217,Khan1984,BHBransden_2001,TTGien_2001} for a considerable time, and 
the latest calculations ~\cite{TTGien_2001} have confirmed the accuracy and provided a valuable benchmark for theoretical approaches.

The complex scaling method (CSM) 
has been widely used to calculate the resonance energies and widths of various few-body systems~\cite{HO19831}
while its application to the non-resonant scattering problems has been relatively limited.
Suzuki et al.~\cite{10.1143/PTP.113.1273,10.1143/PTP.119.949,PhysRevC.89.034322,PhysRevC.104.014325} proposed a CSM calculation of phase shifts by investigating the relationship 
among the complex eigenenergies of the CSM Hamiltonian, continuum level density (CLD), and scattering matrices.
In the CSM calculation of the phase shifts, once the complex eigenenergies of the CSM Hamiltonian are obtained, 
the phase shifts can be calculated as smooth functions of collision energies. 
The CSM basis functions are square integrable functions that damp out within a finite space.
This advantage contrasts with other conventional approaches that utilize an explicit form of the scattering wavefunction and 
require individual calculations for each collision energy.

In this study, we demonstrate the effectiveness of the CSM calculation for low-energy positron scattering off positive ions,
where the collision energy is lower than the first excitation energy of the target atom or ion.
To the best of our knowledge, this CSM approach has not been applied 
to the phase-shift calculations of atomic or molecular systems
and can be used as an alternative approach to atomic and molecular scattering problems, including those involving positrons.

To demonstrate the validity of the CSM, we examine the positron collision with noble gas atoms, 
treating it as a two-body problem using model potentials. 
We then investigate the roles of the eigenenergies responsible for determining the phase-shift behavior of low-energy scattering. 
We consider the empirical fact that
numerical convergence requires a large set of complex eigenenergies even in the case of low-energy scattering.
The majority of the complex eigenenergies have a large real part compared to the small collision energy, which is counterintuitive.
To calculate the phase shift using a finite set of eigenenergies, we propose an effective modification for low-energy scattering.
We apply the CSM to positron scattering off a helium ion (He$^+$), hydrogen atom (H), and lithium ion (Li$^{2+}$) as examples of a three-body problem. We employ the Gaussian expansion method~\cite{GEM2003} and accurately describe the inter-particle interactions. We then perform analogous computations for four-body systems, namely, positron scattering off a He atom and Li$^+$ ion.

The remainder of this paper is organized as follows. In section~\ref{theory}, we provide an overview of the CSM-based calculation of the continuum-level density and phase shift. Subsequently, we describe the basis functions for the calculation of positron-atom scattering.
In section~\ref{RandD}, we highlight a case study of e$^+$-Ne scattering and examine the problems in convergence behavior of low-energy phase shifts. Subsequently, we propose a modification for low-energy scattering phase shifts and demonstrate the modification for e$^+$-X (X=Ne, Ar, Kr, Xe) scattering, e$^+$-X (X=H, He$^+$, Li$^{2+}$) three-body scattering and e$^+$-X (X=He, Li$^{+}$) four-body scattering.
Finally, the conclusions are summarized in section~\ref{conclusion}.

Atomic units (a.u.; $m_e = \hbar = e = 1$) were used throughout this study, except where otherwise mentioned. The Bohr radius is denoted as $a_\mathrm{B}$.

\section{Theory}
\label{theory}

\subsection{Phase-shift calculation with complex scaling method}

The framework of the phase-shift calculation using CSM is presented in Refs. ~\cite{10.1143/PTP.113.1273,10.1143/PTP.119.949,PhysRevC.89.034322,PhysRevC.104.014325}.
Here, we briefly describe the outline of the formulation, which is important for the following discussion. 

We consider a Hamiltonian $H$ consisting of a kinetic energy operator $T$ and a potential energy operator $V$. 
In the CSM, all the coordinates are applied to $\exp(i\theta)$ as follows:
\begin{equation}
    \bm{r}\to\exp(i\theta)\bm{r},
\end{equation}
where $\theta$ is a real number, and $i$ is an imaginary unit.
Using the CSM, the Schr\"{o}dinger equation for the complex-scaled Hamiltonian is written as
\begin{equation}
    H(\theta)\Psi(\theta)=E(\theta)\Psi(\theta),
\end{equation}
where the wavefunction $\Psi(\theta)$ is associated with the eigenenergy $E(\theta)$, which is a complex value because the Hamiltonian is no longer Hermitian. 
If the wavefunction $\Psi(\theta)$ is expanded in terms of a finite number $N_\mathrm{max}$ of $L^2$ integrable basis functions, 
$E(\theta)$ takes discrete energies $E_k(\theta) (k=1,2,3,\cdots N_\mathrm{max})$ 
by a diagonalization of the complex scaled Hamiltonian 
as
\begin{align}
    \langle \tilde{\Psi}_{k'}(\theta) | H(\theta) | \Psi_k(\theta) \rangle = E_k(\theta)\delta_{k'k},
\label{eq.ritz_eigen}
\end{align}
where the bra states with $\tilde{\Psi}_{k}$ denote the bi-orthogonal state of the ket state.
The continuum states then become discrete states, which are referred to as called pseudo-states. 
 
The $E_k(\theta)$ of the bound state does not change with a change in $\theta$. 
In contrast, the $E_k(\theta)$ of the scattering state (pseudo-states) almost rotates by $-2\theta$ in the complex energy plane 
because the kinetic energy operator $T$ of the Hamiltonian is expressed as $\exp(-2i\theta)T$ in the complex scaled Hamiltonian.
The $E_k(\theta)$ of a resonance state approaches a complex resonance energy, and
the real and imaginary parts of which represent the resonance energy and width, respectively.
The eigenfunctions $\{\Psi_k(\theta)\}$ satisfy the completeness relation~\cite{GIRAUD2003115,Giraud_2004} approximately as follows:
\begin{align}
     \sum_{k=1}^{N_\mathrm{max}}{\ket{\Psi_k(\theta)}\bra{ \tilde{\Psi }_k(\theta)}} \approx 1.
\label{eq.completeness_approx}
\end{align}


In this work, we calculate the scattering phase shift $\delta(E)$ based on the continuum level density (CLD)
that is defined as~\cite{SHLOMO199217,Levine}
\begin{align}
\mathcal{D}(E) = -\frac{1}{\pi} \mathrm{Im}\left\{ \mathrm{Tr} \left[ G^+(E)-G^+_0(E)  \right] \right\},
\label{eq.cld_add1}
\end{align}
where $G^+(E)=(E+i\varepsilon-H)^{-1}$ and $G^+_0(E)=(E+i\varepsilon-H_0)^{-1}$ are the full and free Green's functions, respectively.
$H$ denotes the full Hamiltonian of the system and $H_0$ is an asymptotic form of the full Hamiltonian. 

The CLD $\mathcal{D}(E)$ is related to the scattering matrix $S(E)$ as~\cite{Levine,TSANG197543,OSBORN1976119,10.1143/PTP.119.949}
\begin{align}
\mathcal{D}(E) 
=\frac{1}{2\pi}{\rm{Im}}\frac{d}{dE}{\rm{ln}}\,{\rm{det}}\,S(E).
\label{eq.de_se}
\end{align}
When the asymptotic interaction between the collision fragments does not include an explicit Coulomb potential energy operator, 
the $H_0$ only contains kinetic energy operators along with the relative coordinate between the fragments and internal Hamiltonians of the collision fragments.
In this case, the $S(E)$ calculated from $\mathcal{D}(E)$ by Eq.~(\ref{eq.de_se}) corresponds to an amplitude of the outgoing spherical Hankel function.
For the case where the both collision fragments have non-zero charges, we include the asymptotic Coulomb potential energy operator in $H_0$ 
so that the scattering matrix $S(E)$ corresponds to the amplitude of the outgoing spherical Coulomb function.
The $H$ and $H_0$ used in this paper are presented in the following sections, see Eqs.~(\ref{H02bd}), (\ref{H03bd}), and (\ref{H04bd}).
 
For a single-channel (elastic) scattering problem, because ${\rm det}\,S(E)=\exp(2i\delta(E))$, where $\delta(E)$ denotes the phase shift, 
the phase shift is obtained from the CLD as
\begin{align}
\delta(E) = \int_{-\infty}^{E}\mathcal{D}(E')dE'. 
\label{eq.phaseshift_cld}
\end{align}
Using the completeness relation of Eq.~(\ref{eq.completeness_approx}), the $\mathcal{D}(E)$ in Eq.~(\ref{eq.cld_add1}) is rewritten as
\begin{align}
\mathcal{D}(E) 
\approx -\frac{1}{\pi}{\rm{Im}} \Biggr[\sum_{k=1}^{N_\mathrm{max}}\frac{1}{E+i\epsilon-E_k(\theta)}-\sum_{k=1}^{N_\mathrm{max}}\frac{1}{E+i\epsilon-E_{0,k}(\theta)}\Biggr],
\label{eq.phaseshift_greenapprox}
\end{align}
where the summation over $k$ runs for all the eigenenergies, $k=1,\cdots,N_\mathrm{max}$, 
obtained by the full diagonalization of the Hamiltonian with the $N_\mathrm{max}$ basis functions
according to Eq. ~(\ref{eq.ritz_eigen}).
The $\{E_k\}$ and $\{E_{0,k}\}$ are the eigenenergies of $H(\theta)$ and $H_0(\theta)$, respectively, 
and are obtained using the \textit{same} basis functions.
%


In the case of single-channel scattering, 
Eqs.~(\ref{eq.phaseshift_cld}) and (\ref{eq.phaseshift_greenapprox}) provide simple expressions for the phase shift:
\begin{align}
\delta(E_\mathrm{col}) =  \left[ \sum_{k=1}^{N_\mathrm{max}} \delta_k(E_\mathrm{col}) - \sum_{k=1}^{N_\mathrm{max}} \delta_{0,k}(E_\mathrm{col}) \right],
\label{eq.delta_Ecol_org}
\end{align}
where
\begin{align}
\delta_k(E_\mathrm{col})=\mathrm{tan}^{-1} \left( \frac{-\mathrm{Im}E_k(\theta)}{\mathrm{Re}E_k(\theta)-E_\mathrm{col}} \right),
\label{eq.delta_k}
\end{align}
and
\begin{align}
\delta_{0,k}(E_\mathrm{col})=\mathrm{tan}^{-1} \left( \frac{-\mathrm{Im}E_{0,k}(\theta)}{\mathrm{Re}E_{0,k}(\theta)-E_\mathrm{col}} \right).
\label{eq.delta_k0}
\end{align}
It should be noted that, on the complex energy plane, 
$\delta_{(0,)k}(E_\mathrm{col})$ corresponds to the angle of depression of $E_{(0,)k}(\theta)$, as measured from 
the real energy $E_\mathrm{col}$.
Thus, the calculation of the phase shift $\delta(E_\mathrm{col})$ is reduced to the sum of the geometrical angles of the complex eigenenergies on the complex energy plane.

\subsection{Two-body systems: e$^+$-X (X=Ne, Ar, Kr, Xe) }
\label{two-body-theo}

We consider noble gas atom (X = He, Ar, Kr, or Xe) targets as a simple positron scattering problem. 
As atoms have a closed-shell structure and the kinetic energy of the positron is less than the excitation energy,
a two-body model is used in the study. The Hamiltonian $H$ can be written as follows:
\begin{align}
H_\mathrm{2bd} = \frac{\bm{p}^2}{2\mu_\mathrm{e}} + V_\mathrm{e^+X}(r),
\end{align}
where $r$ denotes the distance between e$^+$ and the center of mass of the noble gas atom X, and $\bm{p}$ is the momentum operator along with the vector $\bm{r}$. $\mu_\mathrm{e}$ denotes the reduced mass between e$^+$ and X.
$V_\mathrm{e^+X}(r)$ is a model potential that describes the static Coulomb repulsion and induced dipole polarization as 
\begin{align}
     V_\mathrm{e^+X}(r)=\frac{Z}{r}\sum_i^N A_i e^{-\alpha_i r}
     -\frac{\alpha_\mathrm{d}}{2r^4} \left( 1-e^{-\left({r}/{r_c}\right)^6}  \right).
\end{align}
Here, the static potential parameters $\{A_i\}$ and $\{\alpha_i\}$
are obtained to reproduce the elastic cross-section and are given in Refs. ~\cite{PhysRevA.36.467,Arretche2020} for Ne, Ar, and Xe 
and Ref.~\cite{PhysRevA.36.467} for Kr. 
The induced polarization potential parameters $\alpha_\mathrm{d}$ and $r_c$ are obtained from Ref.~\cite{Arretche2020} for Ne, Ar, and Xe 
and Ref.~\cite{PhysRevA.35.5255} for Kr.
For these systems, we define the reference Hamiltonian $H_0$ using only the kinetic energy operator,
\begin{align}
H_{0,\mathrm{2bd}} = \frac{\bm{p}^2}{2\mu_\mathrm{e}}.
\label{H02bd}
\end{align}

According to Eq.~(\ref{eq.ritz_eigen}), we diagonalize $H_\mathrm{2bd}$ and $H_{0,\mathrm{2bd}}$ using Gaussian basis functions as follows:
\begin{align}
    \psi_{klm} (\mathbf{r}) &= \sum_{i=1}^{n_b} C_i^{(k)} r^l e^{-b_i r^2} Y_{lm}(\hat{\mathbf{r}}),
\end{align}
where the nonlinear Gaussian range parameters $\{b_i; i=1,2,\cdots,n_b\}$ (real values) are selected in accordance with the geometrical progression.  $l$ and $m$ are azimuthal and magnetic quantum numbers, respectively. 
The linear coefficients $\{C_i^{(k)}\}$ are determined using the Rayleigh--Ritz variational method for the complex scaling Schr\"{o}dinger equation 
such that $\psi_{klm}$ satisfies the following condition:
\begin{align}
    \langle \tilde{\psi}_{k' lm}(\theta)  | H_\mathrm{2bd}(\theta) | \psi_{k lm}(\theta) \rangle = E_{k}(\theta)\delta_{k k'}.
\label{sch_complex_e+X}
\end{align}
For the asymptotic Hamiltonian $H_{0,\mathrm{2bd}}$, as in the case of Eq.~(\ref{sch_complex_e+X}), we obtain the eigenenergies $\{E_{0k}\}$.
We typically use $n_b=60$ and $0.01 \leq 1/\sqrt{b_i} \leq 300$ $a_\mathrm{B}$.

\subsection{Three-body systems: e$^+$-X (X=H, He$^+$, Li$^{2+}$) }
\label{three-body-theo}

As a second example of positron scattering, we consider e$^+$-X (X=H, He$^+$, Li$^{2+}$) scattering in a three-body treatment. 
These systems consist of three distinguishable particles, e$^+$, e$^-$, and x$^{Z+}$ (the nucleus of X). 
Assuming an infinite mass of X, the three-body Hamiltonian can be written as
\begin{align}
    H_\mathrm{3bd} = \frac{\bm{p}_\mathrm{e^-}^2+\bm{p}_\mathrm{e^+}^2}{2} 
    -\frac{Z}{r_\mathrm{e^-}}+\frac{Z}{r_\mathrm{e^+}}-\frac{1}{r_\mathrm{e^+e^-}},
\end{align}
where $r_\mathrm{e^-(e^+)}$ denotes the distance between the X and e$^-$(e$^+$), and
$Z$ is the atomic number of the nucleus.
$r_\mathrm{e^+e^-}$ is the distance between e$^-$ and e$^+$. 
For the cases X=He$^+$ and Li$^{2+}$, owing to the repulsive Coulomb interaction remaining in the asymptotic distance, 
the corresponding asymptotic Hamiltonian $H_{0,\mathrm{3bd}}$ is generally expressed as 
\begin{align}
    H_{0,\mathrm{3bd}} = \frac{\bm{p}_\mathrm{e^-}^2+\bm{p}_\mathrm{e^+}^2}{2} 
    -\frac{Z}{r_\mathrm{e^-}}+\frac{Z-1}{r_\mathrm{e^+}},
\label{H03bd}
\end{align}
where the former term of the potential energy operators, $-{Z}/{r_\mathrm{e^-}}$, provides a correct threshold energies of 
collision fragments, and the second term, ${(Z-1)}/{r_\mathrm{e^+}}$, provides the asymptotic repulsive Coulomb interaction 
between the collision fragments for $Z>1$ cases.

\begin{figure}[t]
    \centering
    \resizebox{0.45\textwidth}{!}{\includegraphics[angle=0]{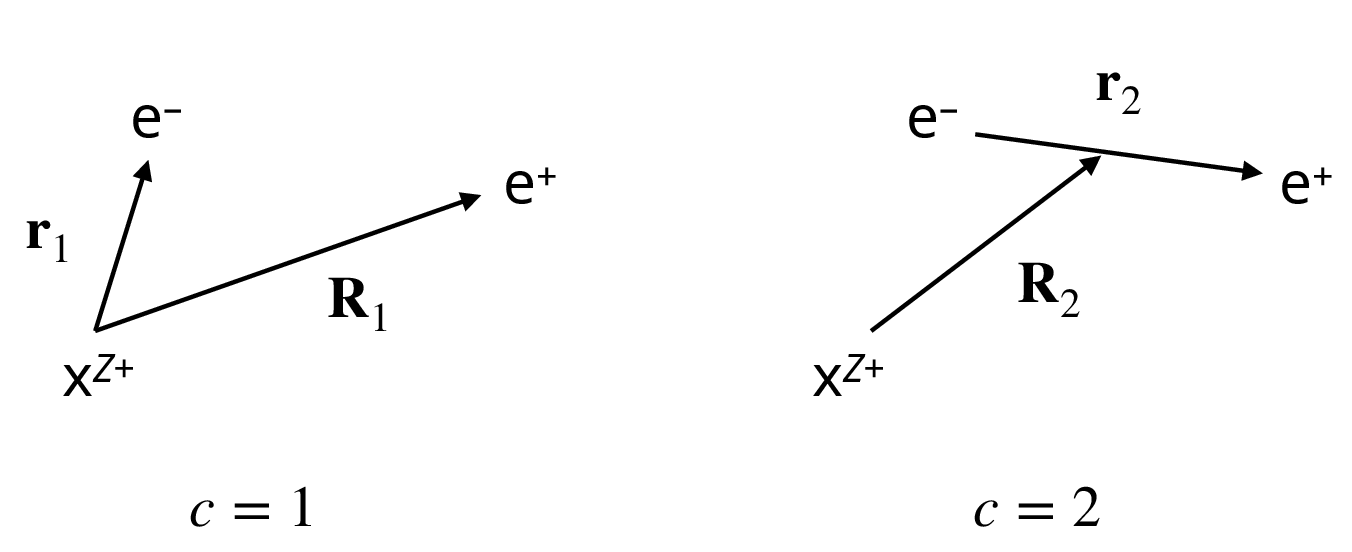}}
    \caption{Two sets of coordinates in e$^+$-H/He$^+$/Li$^{2+}$ scattering.}
    \label{posHepls_coord}
\end{figure}

The wavefunction of the three-body system is written as
\begin{align}
    \Psi_{kJM} &= \sum_{c=1}^{2} \sum_{l,L} \sum_i  
     \left( C_{clLi}^{(k)} \cos{(\beta \nu_{clLi} R_c^2)} +D_{clLi}^{(k)} \sin{(\beta \nu_{clLi} R_c^2)} \right)  
     \nonumber \\ &\times
     r_c^{l}R_c^{L} \exp\left({-\mu_{clLi} r_c^2-\nu_{clLi} R_c^2} \right)
     \left[ Y_{l}(\hat{\bm{r}}_c)\otimes Y_{L}(\hat{\bm{R}}_c) \right]_{JM},
\label{eq:3bdWF}
\end{align}
where the coordinate sets $(\bm{r}_1,\bm{R}_1)=(\bm{r}_\mathrm{e^-},\bm{r}_\mathrm{e^+})$
and $(\bm{r}_2,\bm{R}_2)=(\bm{r}_\mathrm{e^-e^+},\bm{R})$ are presented in Fig.~\ref{posHepls_coord}.
The parameters $\{\nu_{clLi}\}$ and $\{\mu_{clLi}\}$ are selected in accordance with the geometrical progression. 
The typical ranges are $0.01\leq {1}/{\sqrt{\mu_{clLi}}} \leq 20$ $a_\mathrm{B}$ and $0.05\leq {1}/{\sqrt{\nu_{clLi}}} \leq 80$ $a_\mathrm{B}$.
The two spherical harmonics are coupled to provide the total angular momentum $J$ and its projection onto the $z$-axis, $M$.
The parameter $\beta=1.5$ introduces oscillations in the Gaussian functions and increases the orthogonality of the basis functions.
Linear coefficients $C_{clLi}^{(k)}$ and $D_{clLi}^{(k)}$ are determined by the diagonalization as in Eq.~(\ref{eq.ritz_eigen}).
The eigenenergies $\{E_{k}\}$ and $\{E_{0k}\}$ are obtained in a manner similar to that in Eq.~(\ref{sch_complex_e+X}).

\subsection{Four-body systems: e$^+$-X (X=He, Li$^+$) }
\label{four-body-theo}

Similar to section~\ref{three-body-theo}, the four-body Hamiltonian for e$^+$-X (X=He, Li$^+$) scattering is expressed as
\begin{align}
    H_\mathrm{4bd} = &\frac{\bm{p}_\mathrm{e^-_a}^2+\bm{p}_\mathrm{e^-_b}^2+\bm{p}_\mathrm{e^+}^2}{2} 
    -\frac{Z}{r_\mathrm{e^-_a}}-\frac{Z}{r_\mathrm{e^-_b}}+\frac{1}{r_{\mathrm{e^-_a}\mathrm{e^-_b}}}
    \nonumber \\ &+\frac{Z}{r_\mathrm{e^+}}-\frac{1}{r_\mathrm{e^+e^-_b}}-\frac{1}{r_\mathrm{e^+e^-_b}},
\end{align}
where the two electrons are labeled as $\mathrm{e^-_a}$ and $\mathrm{e^-_b}$.
The asymptotic Hamiltonian $H_{0,\mathrm{4bd}}$ is defined assuming
the distances between the positron and the other particles approach the same distances represented by $r_\mathrm{e^+}$, 
i.e.,
\begin{align}
    H_{0,\mathrm{4bd}} = \frac{\bm{p}_\mathrm{e^-_a}^2+\bm{p}_\mathrm{e^-_b}^2+\bm{p}_\mathrm{e^+}^2}{2} 
    -\frac{Z}{r_\mathrm{e^-_a}}-\frac{Z}{r_\mathrm{e^-_b}}+\frac{1}{r_{\mathrm{e^-_a}\mathrm{e^-_b}}}+\frac{Z-2}{r_\mathrm{e^+}}.
\label{H04bd}
\end{align}

\begin{figure}[t]
    \centering
    \resizebox{0.45\textwidth}{!}{\includegraphics[angle=0]{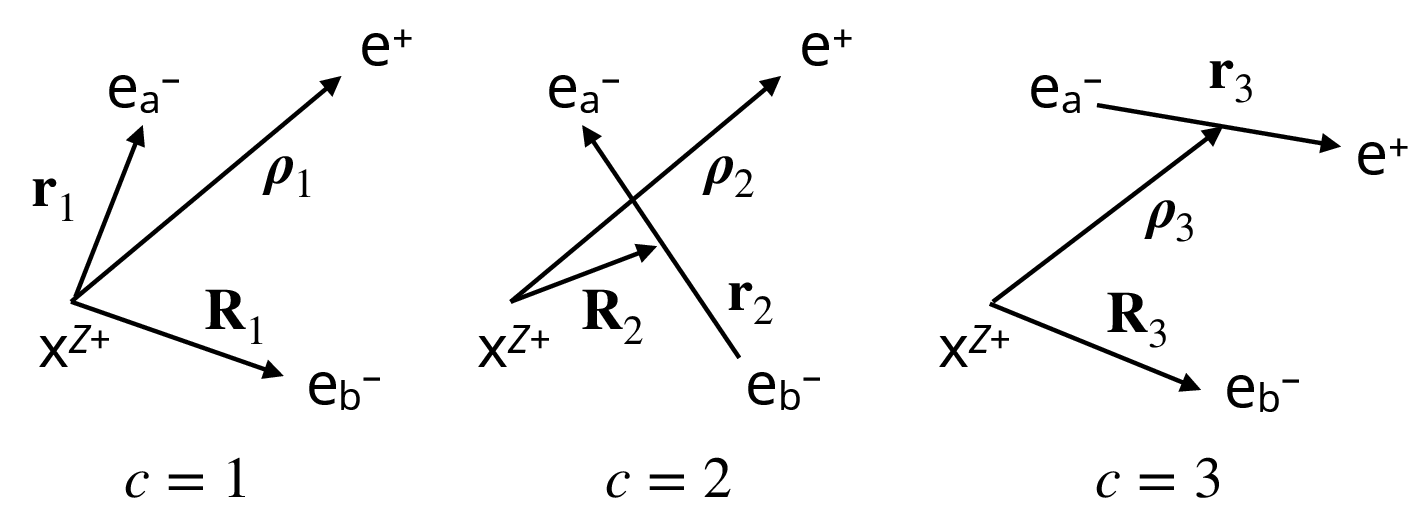}}
    \caption{Three sets of coordinates in e$^+$-He$^+$ scattering.}
    \label{posHe_coord}
\end{figure}

The wavefunction of the four-body system is written as
\begin{align}
    \Psi_{kJM} &= \nonumber\\ &\sum_{c=1}^{3} \sum_{l,L,\lambda} \sum_i  
     \left( C_{clL\lambda i}^{(k)} \cos{(\beta \nu_{clLi} \rho_c^2)} +D_{clL\lambda i}^{(k)} \sin{(\beta \nu_{clLi} \rho_c^2)} \right)  
     \nonumber \\ &\times
     r_c^{l}R_c^{L}\rho_c^{\lambda} \exp\left({-\mu_{clL\lambda i} r_c^2-\nu_{clL\lambda i} R_c^2-\zeta_{clL\lambda i} \rho_c^2} \right) \nonumber \\ 
     &\times
     \left[ \left[ Y_{l}(\hat{\bm{r}}_c)\otimes Y_{L}(\hat{\bm{R}}_c) \right]_\Lambda \otimes Y_{\lambda}(\hat{\bm{\rho}}_c) \right]_{JM} 
+ (\mathrm{a}\leftrightarrow \mathrm{b}),
\label{eq:4bdWF}
\end{align}
where the coordinate sets $(\bm{r}_c,\bm{R}_c,\boldsymbol{\rho}_c)=(\bm{r}_\mathrm{e^-},\bm{r}_\mathrm{e^+})$ are presented 
in Fig.~\ref{posHe_coord}. 
The last term $\mathrm{a}\leftrightarrow \mathrm{b}$ is an electron-permutated basis function because we consider the electronic spin singlet state 
and space-symmetric wavefunctions throughout this paper.  
The non-linear parameters are selected in a manner similar to that of the three-body problem.  
Linear coefficients $C_{clL\lambda i}^{(k)}$ and $D_{clL\lambda i}^{(k)}$ are determined by the diagonalization as in Eq.~(\ref{eq.ritz_eigen}).

\section{Results and Discussion}
\label{RandD}
\subsection{Case-study of potential scattering of e$^+$-Ne collision}
\label{sec.Ne_case}

\begin{figure}[t]
    \centering
    \resizebox{0.48\textwidth}{!}{\includegraphics[angle=0]{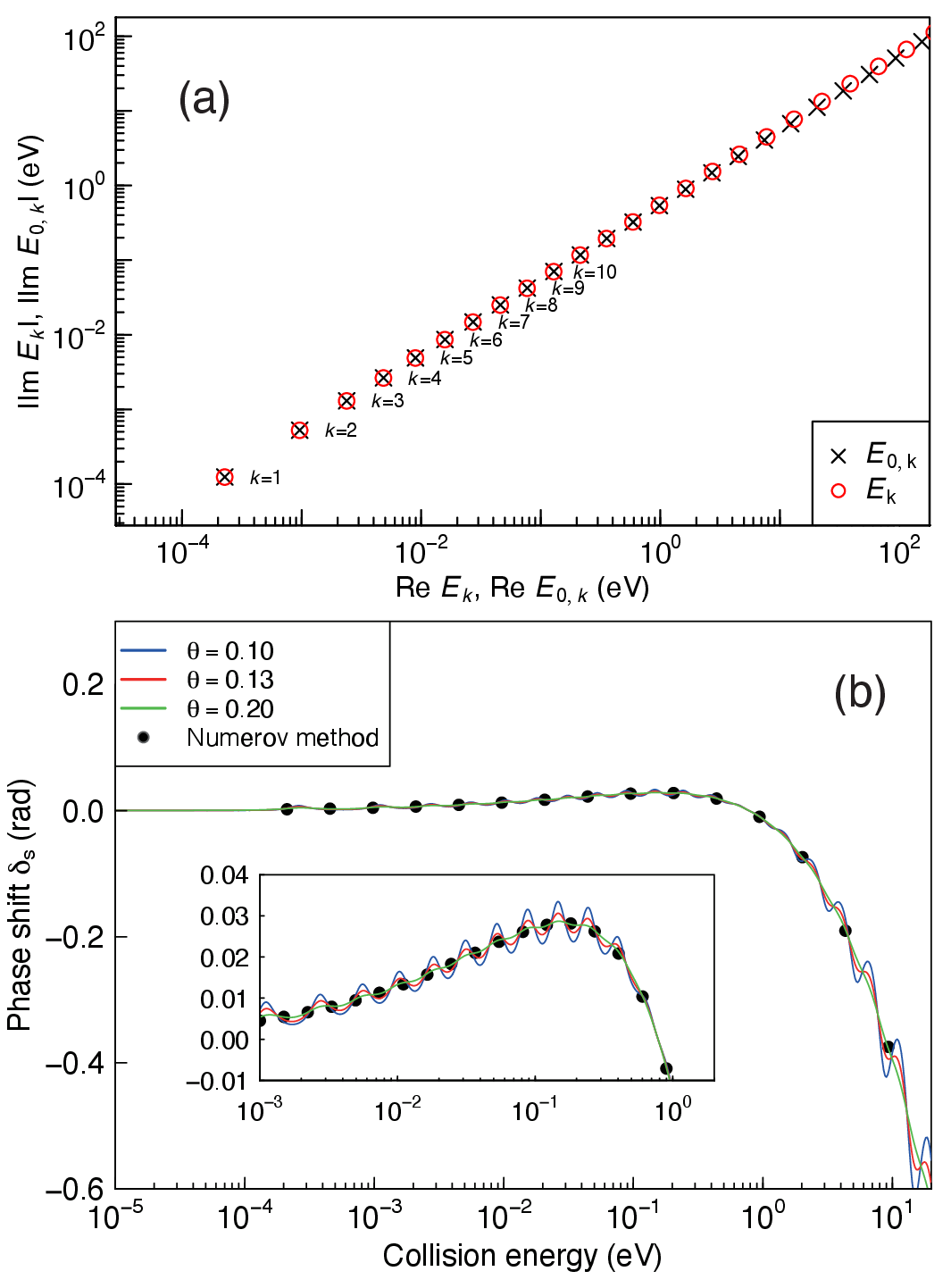}}
    \caption{(a) Complex eigenenergies $\{E_{k}\}$ and $\{E_{0,k}\}$ calculated for $\mathrm{e}^++\mathrm{Ne}$ 
at the complex scaling parameter $\theta=0.25$.
(b) Convergence of the S-wave phase shift of $\mathrm{e}^++\mathrm{Ne}$ scattering against complex scaling parameters $\theta=0.10,0.13,0.25$. 
The phase shifts are compared with those calculated using the Numerov method. 
The inset is a close-up view of the phase shift behavior in the vicinity of 10$^{-2}$-10$^0$ eV.}
    \label{converge}
\end{figure}

We begin the discussion with a case study of e$^+$-Ne scattering.
Using 60 Gaussian basis functions as described in section~\ref{two-body-theo}, 
the 60 complex eigenenergies $\{E_k\}$ and $\{E_{0,k}\}$ are obtained for complex-scaled Hamiltonians $H_\mathrm{2bd}$ and $H_\mathrm{0,2bd}$, respectively.
Figure~\ref{converge}(a) presents the lowest 23 eigenenergies calculated with the complex scaling parameter $\theta=0.25$ for S-wave scattering. 
The system has no bound or resonance states, and all eigenenergies obtained are attributed to continuum states.
Figure~\ref{converge}(b) presents the S-wave phase shifts calculated for several complex scaling angles: $\theta=0.10,0.13$, and $0.25$.
To verify our calculations, we calculated the phase shifts using the Numerov method with high-precision. 
The phase shift realized using a small complex scaling parameter, e.g., $\theta=0.10$, exhibits an oscillation around the exact phase shift obtained using the Numerov method. 
By increasing the scaling angle, the oscillation is reduced and converges to an exact value. 
The calculation of $\theta=0.25$ produces almost the same phase shift as that obtained using the Numerov method 
below the first excitation energy of Ne.
For a small collision energy, the phase shift increases as the collision energy increases and then decreases, thus crossing zero at approximately $E_\mathrm{col}=0.8$ eV.
This behavior results in a minimum S-wave cross-section and a minimum total cross-section, which is known as the Ramsauer--Townsend effect~\cite{Arretche2020}.

\subsection{Problems in convergence behavior of low-energy phase shift}
\label{sec.pairs}

\begin{figure}[t]
    \centering
    \resizebox{0.48\textwidth}{!}{\includegraphics[angle=0]{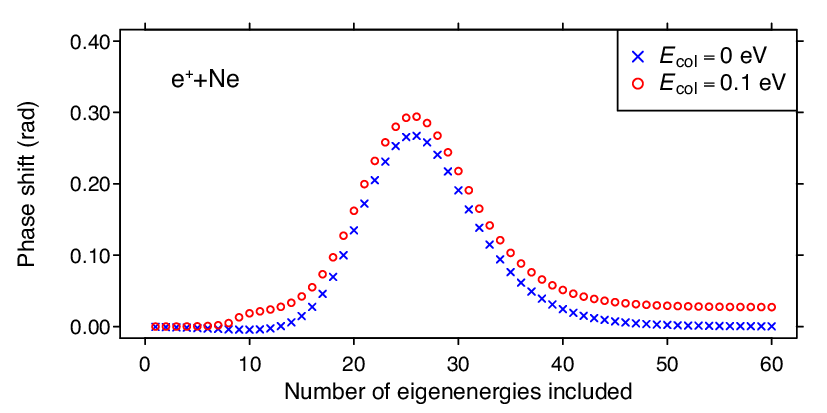}}
    \caption{Convergence of the S-wave phase shift $\delta^{(n)}(E_{\rm col})$ (see Eq.~(\ref{eq.delta_n}) for definition) 
of $\mathrm{e}^+$+Ne at $E_{\rm col}=0$ and 0.1 eV against the number of eigenenergies included $n$.}
    \label{fig.convergence_against_kmax_Ne}
\end{figure}

In section~\ref{sec.Ne_case}, we confirmed that the phase-shift calculation performed using all ($N_\mathrm{max}=60$ in this case) eigenenergies
provides an accurate result for the e$^+$-Ne scattering when the complex scaling angle $\theta$ is sufficiently large.
In this section, we examine the convergence of the phase shift against the number of eigenenergies included.
This consideration is crucial when we calculate three-body and four-body scattering problems.

As observed in Fig.~\ref{converge}(a), the eigenenergies $\{E_{k}\}$ are close to $\{E_{0,k}\}$.
Thus, we can sort these eigenenergies by $\mathrm{Re}\,E_{(0,)k}$ in ascending order and truncate Eq. ~(\ref{eq.delta_Ecol_org}) as
\begin{align}
\label{eq.delta_n}
\delta^{(n)}(E_\mathrm{col}) :=  \sum_{k=1}^{n} \left[ \delta_k(E_\mathrm{col}) - \delta_{0,k}(E_\mathrm{col}) \right],
\end{align}
where $n\leq N_\mathrm{max}$. 
Figure~\ref{fig.convergence_against_kmax_Ne} illustrates the e$^+$+Ne scattering phase shift $\delta^{(n)}(E_\mathrm{col})$ as a function of $n$.
According to Levinson's theorem, the exact phase shift $\delta(E_\mathrm{col}=0\,{\rm eV})$ is zero.
As shown in Section ~\ref{sec.Ne_case}, $\delta^{(n=N_\mathrm{max})}(E_\mathrm{col}=0\,{\rm eV})$ is reproduced to zero. However, 
the convergence behavior of $\delta^{(n)}(E_\mathrm{col})$ is not trivial; $\delta^{(n)}(E_\mathrm{col}=0\,{\rm eV})$ first increases as the number of 
eigenenergies included in ($n$) increases. $\delta^{(n)}(E_\mathrm{col})$ decreases and finally converges to zero.
A similar trend was observed for $\delta^{(n)}(E_\mathrm{col}=0.1\,{\rm eV})$ although the converged phase shift is not zero.
This indicates that we require almost all the eigenenergies having real parts much larger than the collision energy $E_\mathrm{col}$.
For example, in the present case, $\mathrm{Re}\,E_{k=25}\approx 250$ eV, which is counterintuitively larger than $E_\mathrm{col}$.

\begin{figure}[t]
    \centering
    \resizebox{0.47\textwidth}{!}{\includegraphics[angle=0]{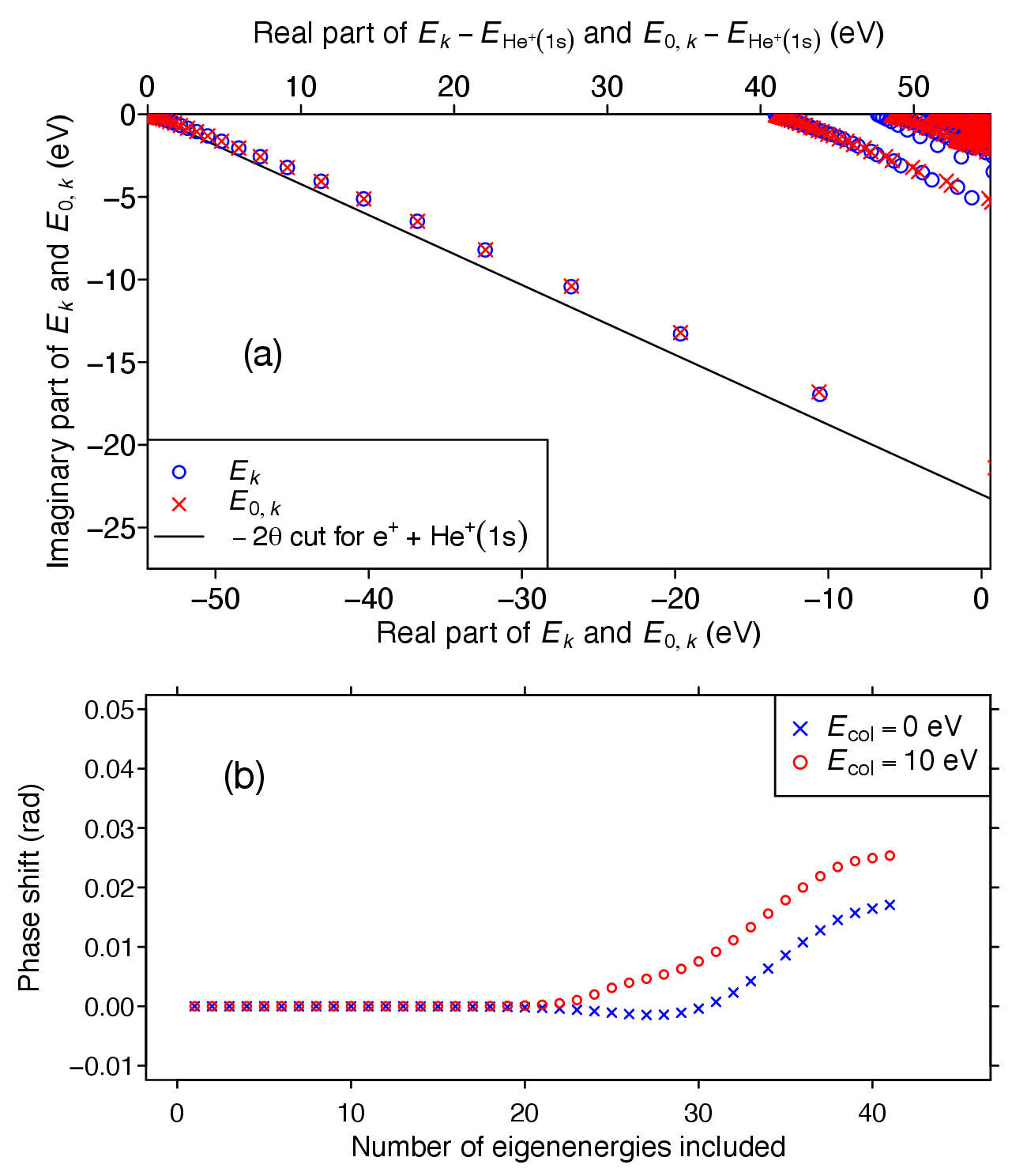}}
    \caption{(a) Complex eigenenergies $\{E_{k}\}$ and $\{E_{0,k}\}$ calculated for $\mathrm{e}^++\mathrm{He}^+$ 
at the complex scaling parameter $\theta=0.20$.
(b) Convergence of the S-wave phase shift $\delta^{(n)}(E_{\rm col})$ of $\mathrm{e}^+$+$\mathrm{He}^+$ at $E_{\rm col}=0$ and 10 eV 
against the number of eigenenergies included $n$.}
    \label{fig.convergence_against_kmax_Heion}
\end{figure}

This ``slow'' convergence against $n$ of the truncated phase shift $\delta^{(n)}(E_\mathrm{col})$ causes a problem in its application to few-body scattering.
We demonstrate the calculation of e$^+$+He$^+$ scattering as a typical case of few-body problems.
Figure~\ref{fig.convergence_against_kmax_Heion}(a) presents the complex eigenenergies of a three-body system ($\mathrm{e}^+$, $\mathrm{e}^-$, and He$^{2+}$), 
which are obtained by diagonalization with approximately 4500 basis functions.
Because of the internal energy of He$^+(1s)$ $E_\mathrm{He^+(1s)}=-54.4$ eV, the lowest $\{E_k\}$ and $\{E_{0,k}\}$ values are located just above the threshold energy $-54.4$ eV. 
In contrast to the two-body problem of e$^+$+Ne scattering,
the eigenenergies at high energies cannot be attributed to any of the specific physical channels because 
there are several inelastic channels, and the corresponding eigenfunction must be a multi-channel scattering wavefunction. 
In the present case, we observe a series of eigenenergies that rotate by $2\theta$ 
around the lowest threshold energy e$^+$+He$^+(1s)$.
Taking these complex eigenenergies into consideration, we examine the convergence of $\delta^{(n)}(E_\mathrm{col})$, as shown in Fig. ~\ref{fig.convergence_against_kmax_Heion}(b).
Although we included eigenenergies having real parts exceeding 40 eV, $\delta^{(n)}(E_\mathrm{col}=0\,{\rm eV})$ and 
$\delta^{(n)}(E_\mathrm{col}=10\,{\rm eV})$ do not reach convergence. 
At least $\delta^{(n)}(E_\mathrm{col}=0\,{\rm eV})$ should have converged to zero, while $\delta^{(n=40)}(E_\mathrm{col}=0\,{\rm eV})$ deviated significantly from zero. 

\subsection{Modification to the low-energy phase shift}
\label{sec:modification}

\begin{figure}[t]
    \centering
    \resizebox{0.48\textwidth}{!}{\includegraphics[angle=0]{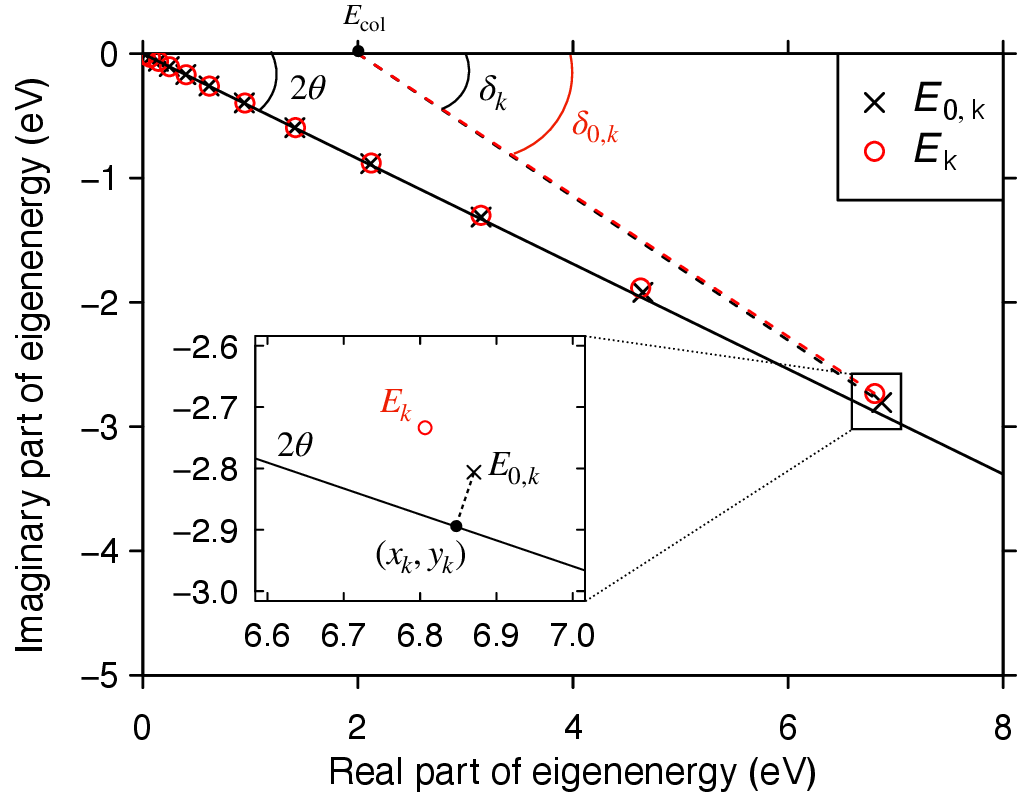}}
    \caption{Schematic of complex eigenenergies and the related angles of depression on the complex energy plane. 
$(x_k,y_k)$ is selected as the closest point to $E_{0,k}$ on the $2\theta$ line.}
    \label{fig.illust_pairs}
\end{figure}

In this subsection, we discuss the effective modification of the low-energy scattering phase shift when the number of eigenenergies is limited.
As observed in the convergence behavior of $\delta^{(n)}(E_\mathrm{col})$, the high-lying eigenenergies contributed to the determination of the low-energy phase shift.
For practical purposes in atomic physics, it is worth proposing an effective modification to calculate low-energy phase shift even 
when only a limited number of eigenenergies close to the $E_\mathrm{col}$ are obtained.
The phase shift is obtained  by summing the angles of depression of $E_{(0,)k}$ measured from $E_\mathrm{col}$; therefore,
the contributions from the high-lying eigenenergies are expected to be insensitive to changes in $E_\mathrm{col}$ in the low-energy scattering region. 

Thus, we formulate the contributions from the high-lying eigenenergies to the phase shift, with particular emphasis on the $E_\mathrm{col}$ dependency.
We consider a situation wherein the eigenenergies $\{E_k\}$ are observed close to $\{E_{0,k}\}$ and can define eigenenergy pairs $(E_k, E_{0,k})$ 
as illustrated in Fig.~\ref{fig.illust_pairs}.
Each eigenenergy pair causes the following angular difference,
\begin{align}
\delta_{\mathrm{diff},k}(E_\mathrm{col}):=\delta_k(E_\mathrm{col}) - \delta_{0,k}(E_\mathrm{col}).
\label{eq.delta_diff}
\end{align}
The sum of which over $k$ yields a phase shift $\delta(E_\mathrm{col})$.
It should be noted that, in Fig.~\ref{fig.illust_pairs}, $\{E_{0,k}\}$ are not exactly on the $2\theta$ cut owing to the asymptotic Coulomb potential operator contained in $H_0$
and/or the numerical accuracy, depending on the number and types of basis functions. 

We define a point $(x_k,y_k)$ on $2\theta$ cut as $y_k=-\tan(2\theta)x_k$
such that the distance between $(x_k,y_k)$ and $E_{0,k}$ is minimized (see the inset of Fig. ~\ref{fig.illust_pairs}).
Then, $E_k$ and $E_{0,k}$ are expressed as deviations from the same point $(x_k,y_k)$ as follows: 
\begin{align}
\label{eq.delta_xy}
\left(\mathrm{Re}\,E_k,\mathrm{Im}\,E_k\right) =: \left(x_k(1+\Delta_{x,k}),y_k(1+\Delta_{y,k})\right),
\end{align}
and
\begin{align}
\label{eq.delta_xy0}
\left(\mathrm{Re}\,E_{0,k},\mathrm{Im}\,E_{0,k}\right) = :\left(x_k(1+\Delta_{0x,k}),y_k(1+\Delta_{0y,k})\right),
\end{align}
where $\Delta_{x,k}$, $\Delta_{y,k}$, $\Delta_{0x,k}$, and $\Delta_{0y,k}$ are dimensionless real numbers. 
Using $\Delta_{x,k}$ and $\Delta_{y,k}$, the arctangent argument in Eq.~(\ref{eq.delta_k}) can be written as
\begin{align}
\frac{-\mathrm{Im}E_k(\theta)}{\mathrm{Re}E_k(\theta)-E_\mathrm{col}} 
&=\tan{2\theta} \frac{1+\Delta_{y,k}}{1+{\Delta}_{x,k}-E_\mathrm{col}/x_k},
\label{eq.atan-argument}
\end{align}
and a similar expression can be obtained in Eq. ~(\ref{eq.delta_xy0}).

We now consider the case wherein ${\Delta}_{x,k},\Delta_{y,k}\ll 1$,
and $E_\mathrm{col}\ll x_k$, i.e., the eigenenergy pairs that are located in a significantly high-energy region compared with $E_\mathrm{col}$.
For brevity, we rewrite
\begin{equation}
\tilde{\Delta}_{x,k} := \Delta_{x,k}-E_\mathrm{col}/x_k.
\end{equation}
On expanding $(1+\tilde{\Delta}_{x,k})^{-1}$ in Eq.~(\ref{eq.atan-argument}) 
and then expanding Eq.~(\ref{eq.delta_k}), $\delta_k(E_\mathrm{col})$ can be written as a perturbation expansion as (see Appendix for details) 
\begin{align}
\delta_k(E_\mathrm{col})
&= 2\theta +\frac{1}{2}\mathrm{sin}(4\theta)\left\{-\tilde{\Delta}_{x,k}+\Delta_{y,k} +\mathrm{cos}^2(2\theta)\tilde{\Delta}^2_{x,k} \right. \nonumber \nonumber\\
&\hspace{0.8cm}\left.-\mathrm{cos}(4\theta)\tilde{\Delta}_{x,k}\Delta_{y,k}
-\mathrm{sin}^2(2\theta)\Delta^2_{y,k}\right\} + O_3,
\label{eq.delta_expansion}
\end{align}
where $O_3$ denotes the third-order term of the deviations, $\tilde{\Delta}^3_{x,k},\tilde{\Delta}^2_{x,k}\Delta_{y,k},\tilde{\Delta}_{x,k}\Delta^2_{y,k}$, and $\Delta^3_{y,k}$.
Similar expansions can be observed in Eq. ~(\ref{eq.delta_k0}).
Therefore, $\delta_{\mathrm{diff},k}(E_\mathrm{col})$ in Eq. ~(\ref{eq.delta_diff}) can be expressed as
\begin{align}
\delta_{\mathrm{diff},k}(E_\mathrm{col})
&= \frac{1}{2}\mathrm{sin}(4\theta)\left\{-\tilde{\Delta}_{x,k}+\tilde{\Delta}_{0x,k}+\Delta_{y,k}-\Delta_{0y,k}\right. \nonumber\\
&+\mathrm{cos}^2(2\theta)\left(\tilde{\Delta}^2_{x,k}-\tilde{\Delta}^{2}_{0x,k}\right) 
 -\mathrm{sin}^2(2\theta)\left(\Delta^2_{y,k}-\Delta^{2}_{0y,k}\right)  \nonumber\\
&\left.-\mathrm{cos}(4\theta)\left(\tilde{\Delta}_{x,k}\Delta_{y,k}-\tilde{\Delta}_{0x,k}\Delta_{0y,k}\right) \right\} + O_3,
\label{eq.delta_k_dif_approx_Delta}
\end{align}
where $\tilde{\Delta}_{0x(y),k}:=\Delta_{0x(y),k}-E_\mathrm{col}/x_k$.

Because the collision-energy dependence of $\delta_{\mathrm{diff},k}(E_\mathrm{col})$ emerges from $E_\mathrm{col}/x_k$, which appears for every $\tilde{\Delta}_{x(y),k}$ and $\tilde{\Delta}_{0x(y),k}$,
the first-order terms in Eq.~(\ref{eq.delta_k_dif_approx_Delta}) only contains the term of $E_\mathrm{col}^0$, 
and the second-order term contains only the term $E_\mathrm{col}^0$ and $E_\mathrm{col}^1$.
In summary, $\delta_{\mathrm{diff},k}(E_\mathrm{col})$ can be written as
\begin{align}
\delta_{\mathrm{diff},k}(E_\mathrm{col})=a_k+\frac{b_k}{x_k}E_\mathrm{col} + O_3,
\label{eq.delta_k_dif_ak_bk}
\end{align}
where $a_k$ and $b_k$ are independent of $E_\mathrm{col}$ and depend only on the complex scaling angle $\theta$ and the parameters of the eigenenergy pairs $\Delta_{x,k},\Delta_{0x,k},\Delta_{y,k}$, and $\Delta_{0y,k}$.
From Eq.~(\ref{eq.delta_k_dif_ak_bk}), the second term can be neglected for the eigenenergy pairs with $|x_k|\gg E_\mathrm{col}$.
As $x_x\approx \mathrm{Re}\,E_{0,k}$, 
$\delta_{\mathrm{diff},k}(E_\mathrm{col})$ can be considered a \textit{constant} for a large value of $\mathrm{Re}\,E_{0,k}$.
We then approximate the sum of $\delta_{\mathrm{diff},k}(E_\mathrm{col})$ 
originating from the high-lying (large $k$) complex eigenenergies as constants,
the phase shift of the collision energy $E_\mathrm{col}$ can be expressed as
\begin{align}
\delta(E_\mathrm{col}) \approx \sum^{n}_{k=1}\delta_{\mathrm{diff},k} (E_\mathrm{col}) + \delta_{>n,\mathrm{const}},
\label{eq.delta_k_dif_approx}
\end{align}

Although $\delta_{>n,\mathrm{const}}$ is not known in advance, Eq.~(\ref{eq.delta_k_dif_approx}) is valid for $E_\mathrm{col}=0$
and Levinson's theorem for a system with no bound state
\begin{align}
\delta(E_\mathrm{col}=0)=0,
\label{eq.delta_k_dif_approx-1}
\end{align}
provides a constraint.
On subtracting Eq.~(\ref{eq.delta_k_dif_approx}) by that of $E_\mathrm{col}=0$, we obtain
\begin{align}
\delta(E_\mathrm{col}) 
&\approx \sum^{n}_{k=1}\left[ \delta_{\mathrm{diff},k} (E_\mathrm{col}) -\delta_{\mathrm{diff},k} (E_\mathrm{col}=0) \right]
\nonumber \\
&= \delta^{(n)} (E_\mathrm{col}) -\delta^{(n)}(E_\mathrm{col}=0).
\label{eq.delta_k_dif_approx-2}
\end{align}
Thus, even when we obtain a limited number of eigenenergy pairs $n<N_\mathrm{max}$, the subtraction of $\delta^{(n)}(E_\mathrm{col}=0)$ from $\delta^{(n)} (E_\mathrm{col})$ is
expected to improve the accuracy of the phase-shift determination.
Hereinafter, we refer to the modification of Eq. ~(\ref{eq.delta_k_dif_approx-2}) as the ``calibration.''
We apply Eq.~(\ref{eq.delta_k_dif_approx-2}) to several positron-scattering problems in the following subsections.

It would be practically useful to show a way 
estimating the applicable energy range of the calibration by Eq.~(\ref{eq.delta_k_dif_approx-2}).
The $\delta_{\mathrm{diff},k} (E_\mathrm{col})$, the $|a_k x_x/b_k|$ of which satisfies $E_\mathrm{col}\ll |a_k x_x/b_k|$ 
can be considered as a constant with respect to changes in $E_\mathrm{col}$ under low-energy scattering.
The truncation number $n$ should be sufficiently large, such that $\mathrm{Re}\,E_{(0,)n}$ covers $E_\mathrm{col}$. 
Because $(x_k,y_k)$ is selected as the point closest to $E_{0,k}$ on the $2\theta$-cut, 
$x_k$ can be approximated as $\mathrm{Re}\,E_{0,k}$. Thus, by replacing $x_k$ with $\mathrm{Re}\,E_{0,k}$, we obtain
\begin{align}
&\bigg|\frac{a_k \mathrm{Re}\,E_{0,k}}{b_k} \bigg| \nonumber \\ 
&\approx \left|\frac{\left\{ \left(\Delta_{x,k}-\Delta_{0x,k}\right)-\left(\Delta_{y,k}-\Delta_{0y,k}\right)\right\} \mathrm{Re}E_{0,k}}
             { 2\mathrm{cos}^2(2\theta)\left(\Delta_{x,k}-\Delta_{0x,k}\right) -\mathrm{cos}(4\theta)\left(\Delta_{y,k}-\Delta_{0y,k}\right)}\right|.
  \label{eq.ak_bk_estim}
\end{align}
Here, the second-order term in $a_k$ is ignored.
When the complex scaling angle $\theta$ is small (typically $\theta\sim \pi/12$) because
$2\mathrm{cos}^2(2\theta)\sim 2(1-2\theta)^2 \sim 2(1-4\theta)$ and $\mathrm{cos}(4\theta)\sim (1-4\theta)$,
we obtain the following approximate estimation:
\begin{align}
\bigg|\frac{a_k \mathrm{Re}\,E_{0,k}}{b_k} \bigg| \gtrsim \frac{\mathrm{Re}\,E_{0,k}}{2}. 
\end{align} 
Thus, the applicable energy range for Eq.~(\ref{eq.delta_k_dif_approx-2}) is approximately half that of $\mathrm{Re}\,E_{0,n}$.

\subsection{Validity of calibration in $\mathrm{e}^+$-noble gas atom scattering}

In this subsection, we examine the formulations described in the previous subsection for $\mathrm{e}^+$+Ne potential scattering and
demonstrate the validity of the calibration modification.

\begin{figure}[t]
    \centering
    \resizebox{0.48\textwidth}{!}{\includegraphics[angle=0]{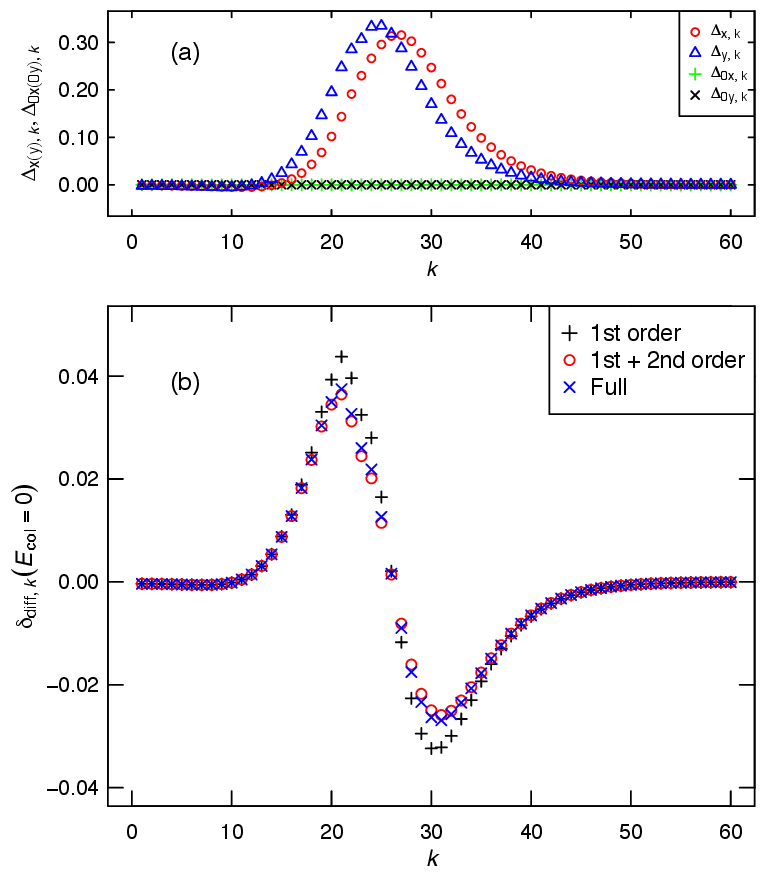}}
    \caption{(a) $\Delta_{x,k}$, $\Delta_{y,k}$, $\Delta_{0x,k}$, and $\Delta_{0y,k}$ of eigenenergies of $\mathrm{e}^+$+Ne scattering 
corresponding to Fig.~\ref{fig.decomposed}. 
(b) Reproducibility of $\delta_{{\rm diff},k}$ (at $E_{\mathrm{col}} = 0$) by the first- and second-order terms calculated 
from $\Delta_{x,k}$, $\Delta_{y,k}$, $\Delta_{0x,k}$, and $\Delta_{0y,k}$
according to Eq.~(\ref{eq.delta_k_dif_approx_Delta}).} 
    \label{fig.deltas}
\end{figure}

We first investigate the influence of each eigenenergy pair on the phase shifts and examine the validity of Eq. ~(\ref{eq.delta_k_dif_ak_bk}).
We use the same calculation for $\mathrm{e}^+$+Ne potential scattering presented in subsection~\ref{sec.Ne_case}, i.e., $\theta=0.25$.
Figure~\ref{fig.deltas}(a) shows that $\Delta_{x,k}$, $\Delta_{y,k}$, $\Delta_{0x,k}$, and $\Delta_{0y,k}$ 
of $\mathrm{e}^+$+Ne potential scattering at $E_\mathrm{col}=0$ eV.
$|\Delta_{x,k}|$ and $|\Delta_{y,k}|$ are in the range of $10^{-4}-10^{0}$, while $|\Delta_{0x,k}|$ and $|\Delta_{0y,k}|$ are smaller than $10^{-10}$.
It is convincing that small $|\Delta_{0x,k}|$ and $|\Delta_{0y,k}|$ because the $\{E_{0,k}\}$, eigenenergies of $H_0$ (containing only the kinetic energy operator),
are on the $2\theta$ line. In contrast, $\{E_{k}\}$ deviates from the $2\theta$ line owing to the presence of the potential energy operator $V_\mathrm{e^+Ne}$.

In Fig.\ref{fig.deltas}(b), we examine Eq.~(\ref{eq.delta_k_dif_approx_Delta}), calculated using $\Delta_{(0)x,k}$ and $\Delta_{(0)y,k}$.
The first-order term and 1st+2nd order term of $\delta_{\mathrm{diff},k}$ are compared with the exact $\delta_{\mathrm{diff},k}(E_\mathrm{col}=0)$.
The first-order perturbation makes a dominant contribution to the exact $\delta_{\mathrm{diff},k}$. 
Including up to the second-order terms, the exact $\delta_{\mathrm{diff},k}$ is almost entirely explained.
As the first-order term does not depend on $E_\mathrm{col}$, it is convincing that the gross structure of $\delta_{\mathrm{diff},k}$ does not show collision-energy dependence. 
Thus, the expansion of Eq.(\ref{eq.delta_k_dif_approx_Delta}) is reasonable, 
and according to Eq.~(\ref{eq.delta_k_dif_ak_bk}), for $\delta_{\mathrm{diff},k}$, the high-energy eigenenergy pairs can be considered as constants
compared to $E_{\rm col}$ for low-energy scattering.

\begin{figure}[t]
    \centering
    \resizebox{0.47\textwidth}{!}{\includegraphics[angle=0]{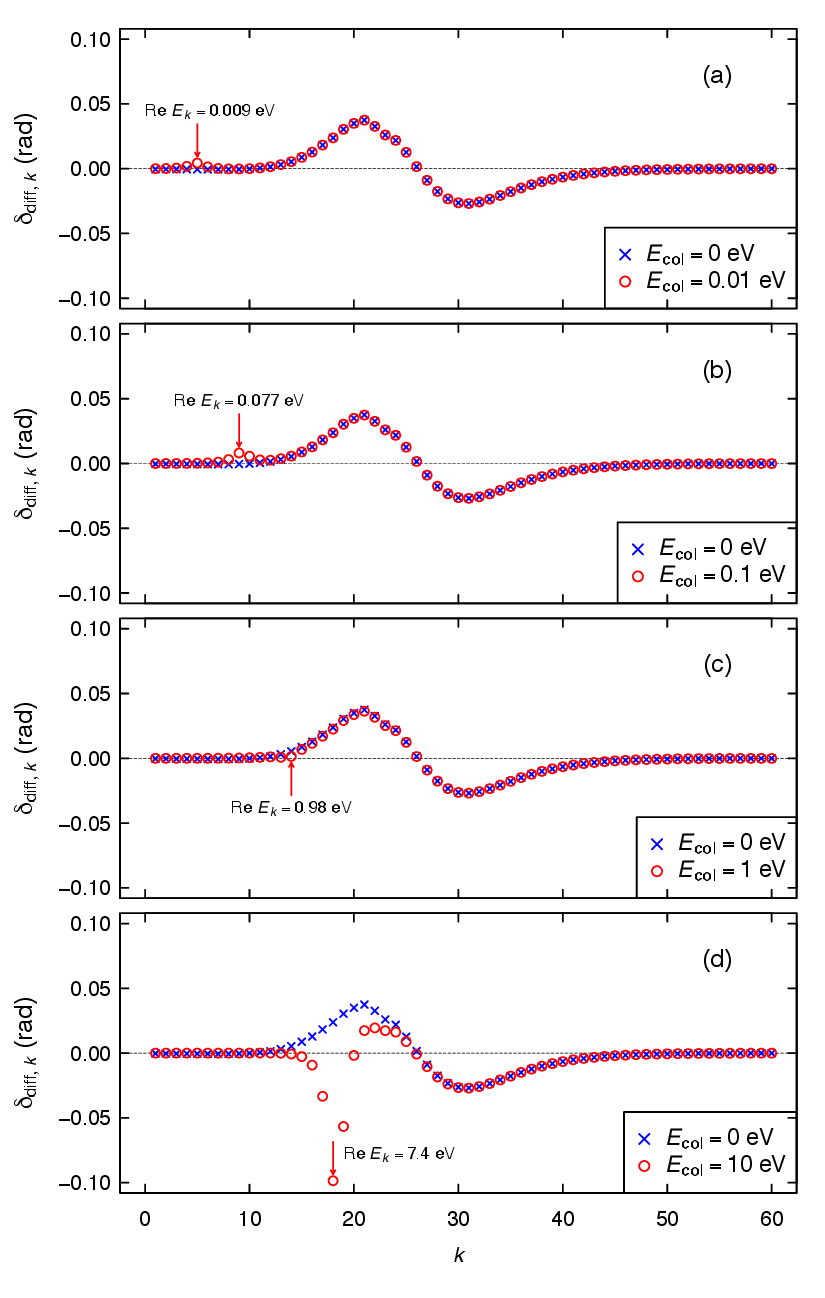}}
    \caption{ S-wave phase shift of $\mathrm{e}^++\mathrm{Ne}$ decomposed into $\delta_{{\rm diff},k}$ (see Eq.~(\ref{eq.delta_diff}) for definition) 
for each pair of the complex eigenenergies. $E_{\rm col}=0.01$, $0.1$, $1$, and $10$ eV are compared with $E_{\rm col}=0$ eV.} 
    \label{fig.decomposed}
\end{figure}

Figure~\ref{fig.decomposed} presents a comparison of $\delta_{\mathrm{diff},k}$ 
of $\mathrm{e}^+$+Ne potential scattering at $E_\mathrm{col}=0$ eV with several collision energies $E_\mathrm{col}=0.01,0.1,1$, and $10$ eV.
The largest $|\delta_{\mathrm{diff},k}|$ for the $E_\mathrm{col}=0$ eV collision is the eigenenergy pair 
near $k=20$ and $30$.
It should be noted that these positive and negative values of $|\delta_{\mathrm{diff},k}|$ totally cancel out 
and result in zero. 
The phase shift for $E_\mathrm{col}>0$ is equivalent to the sum of the differences between $\delta_{\mathrm{diff},k}(E_\mathrm{col})$ and $\delta_{\mathrm{diff},k}(E_\mathrm{col}=0)$.
For each $\delta_{\mathrm{diff},k}(E_\mathrm{col})$, a deviation of $\delta_{\mathrm{diff},k}(E_\mathrm{col})$ from $E_\mathrm{col}=0$ can be observed 
around the $E_\mathrm{col}$, as expected.
For example, the remarkable difference of $\delta_{\mathrm{diff},k}(E_\mathrm{col})$ between $E_\mathrm{col}=0$ eV and $E_\mathrm{col}=0.01$ eV can be observed 
at $\mathrm{Re}\,E_k=0.009$ eV, which is close to $E_\mathrm{col}$. This is also true for $E_\mathrm{col}=0.1$ and $E_\mathrm{col}=10$ eV.
$\delta_{\mathrm{diff},k}$ for $E_\mathrm{col}=1$ eV is very similar to that for $E_\mathrm{col}=0$, 
which is because $E_\mathrm{col}=1$ eV is close to the Ramsauer--Townsend minimum, where the phase shift becomes zero.
In summary, the energy dependence of the phase shift originates from the eigenenergy pairs near the collision energy.
The equation~(\ref{eq.delta_Ecol_org}) formally requires all the eigenenergy pairs, while the eigenenergy pairs
far from the collision energy have the same contribution as $E_\mathrm{col}=0$. Thus, Fig.~\ref{fig.decomposed} clearly
demonstrates the validity of the modification in Eq.~(\ref{eq.delta_k_dif_approx-2}).

\begin{figure}[t]
    \centering
    \resizebox{0.48\textwidth}{!}{\includegraphics[angle=0]{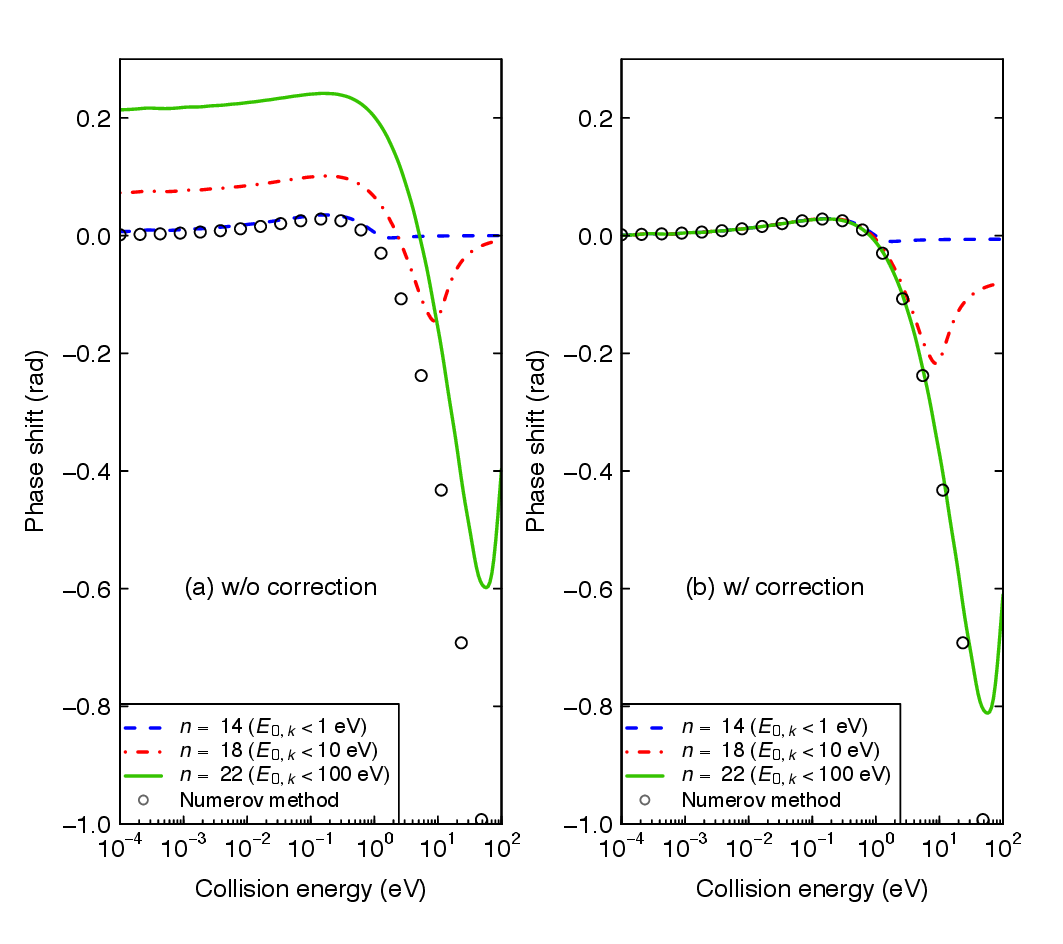}}
    \caption{S-wave phase shifts of $\mathrm{e}^++\mathrm{Ne}$ calculated using a limited number of pairs of complex eigenenergies $n$.
(a) Without calibration, (b) with calibration according to Eq.~(\ref{eq.delta_k_dif_approx-2}).
The phase shifts calculated using the CSM are compared with those calculated using the Numerov method.}
    \label{some_n}
\end{figure}

Figure~\ref{some_n} presents the $\mathrm{e}^+$+Ne S-wave scattering phase shifts obtained using only the lowest 10, 20, and 25 eigenenergy pairs of the 60 pairs in total.
As shown in panel (a), the use of a limited number of eigenenergy pairs results in a non-zero phase shift at $E_\mathrm{col}=0$. 
However, the ``calibrated'' phase shifts 
reproduced the exact phase-shift behavior of low-energy collisions for $n=14, 18,$ and $22$.
As the number of eigenenergy pairs increased, the collision energy range, where the ``calibrated'' phase shifts reproduce the exact phase shifts, becomes wider. 

\begin{figure}[t]
    \centering
    \resizebox{0.48\textwidth}{!}{\includegraphics[angle=0]{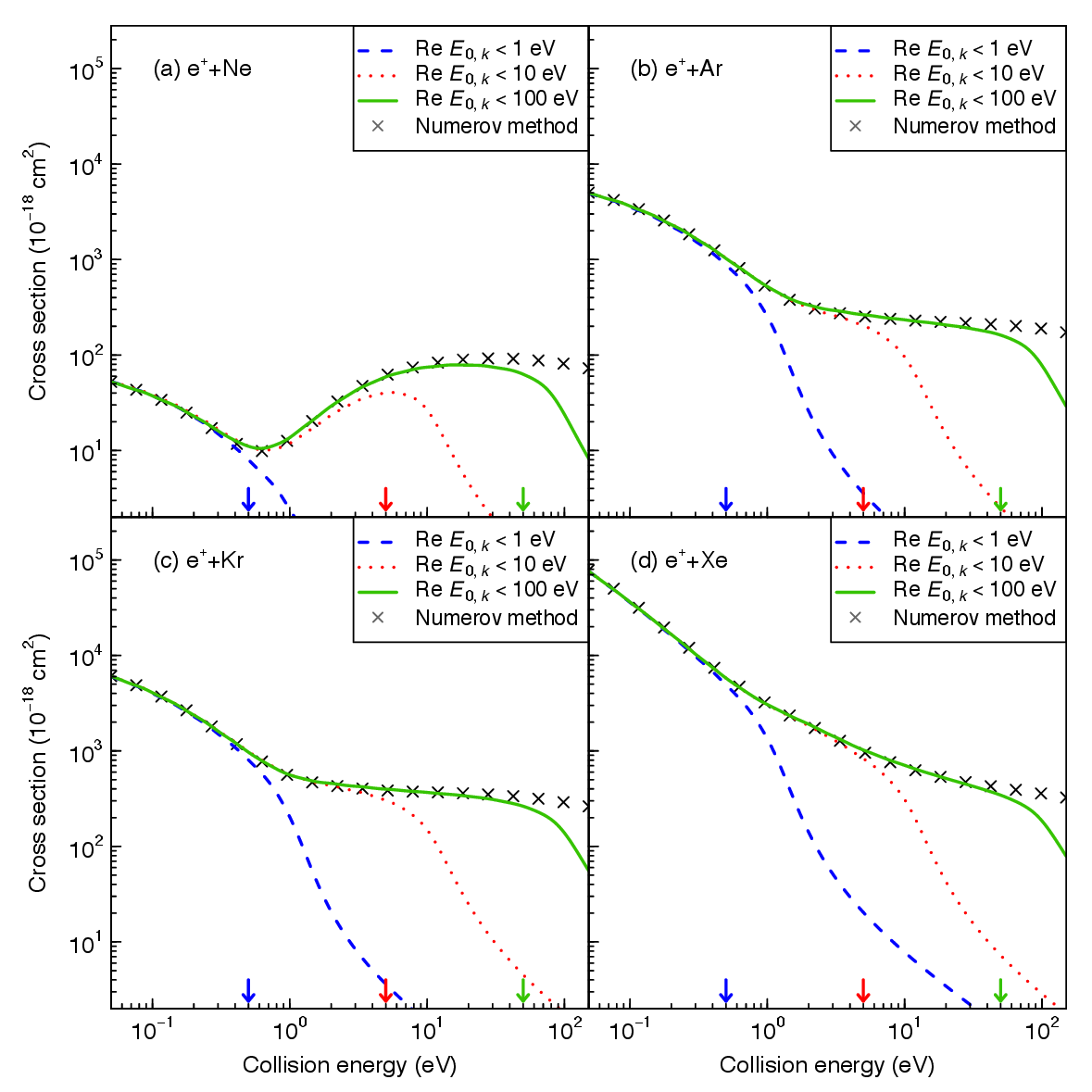}}
    \caption{Elastic scattering cross sections of $\mathrm{e}^+$+X (X=Ne, Ar, Kr, and Xe) calculated using the calibrated phase shifts are 
presented against the collision energy.
The arrows of colors matching the corresponding line indicate the $\mathrm{Re}\,E_{0,n}/2$ to show the expected applicable range.
}
    \label{cross-section-all}
\end{figure}

Using this modification, the elastic cross sections of $\mathrm{e}^++\mathrm{X} \;(\mathrm{X}=\mathrm{Ne},\mathrm{Ar},\mathrm{Kr},\mathrm{Xe})$ 
are calculated as shown in Fig.~\ref{cross-section-all}.
In these calculations, a pair of eigenenergies that satisfies $\mathrm{Re}\ E_{(0,)k} \le E_\mathrm{cut}$ are used, where $E_\mathrm{cut}=1,100,$and $1000$ eV.
The partial waves are considered from the S-wave ($l=0$) to $l=8$ for Ne, Ar, and Kr, and to $l=11$ for Xe such that the cross-sections converge against the partial waves.
With an increase in the number of eigenenergy pairs used for the calculation, the total elastic cross sections reproduced the exact ones calculated using the Numerov method. 
The indicator $\mathrm{Re}\ E_{0,k_\mathrm{max}}/2$ also agrees well with the maximum energy at which the modification is valid.

\subsection{Three- and four-body calculations of e$^+$ scattering using limited number of eigenenergies}

\begin{figure}[t]
    \centering
    \resizebox{0.45\textwidth}{!}{\includegraphics[angle=0]{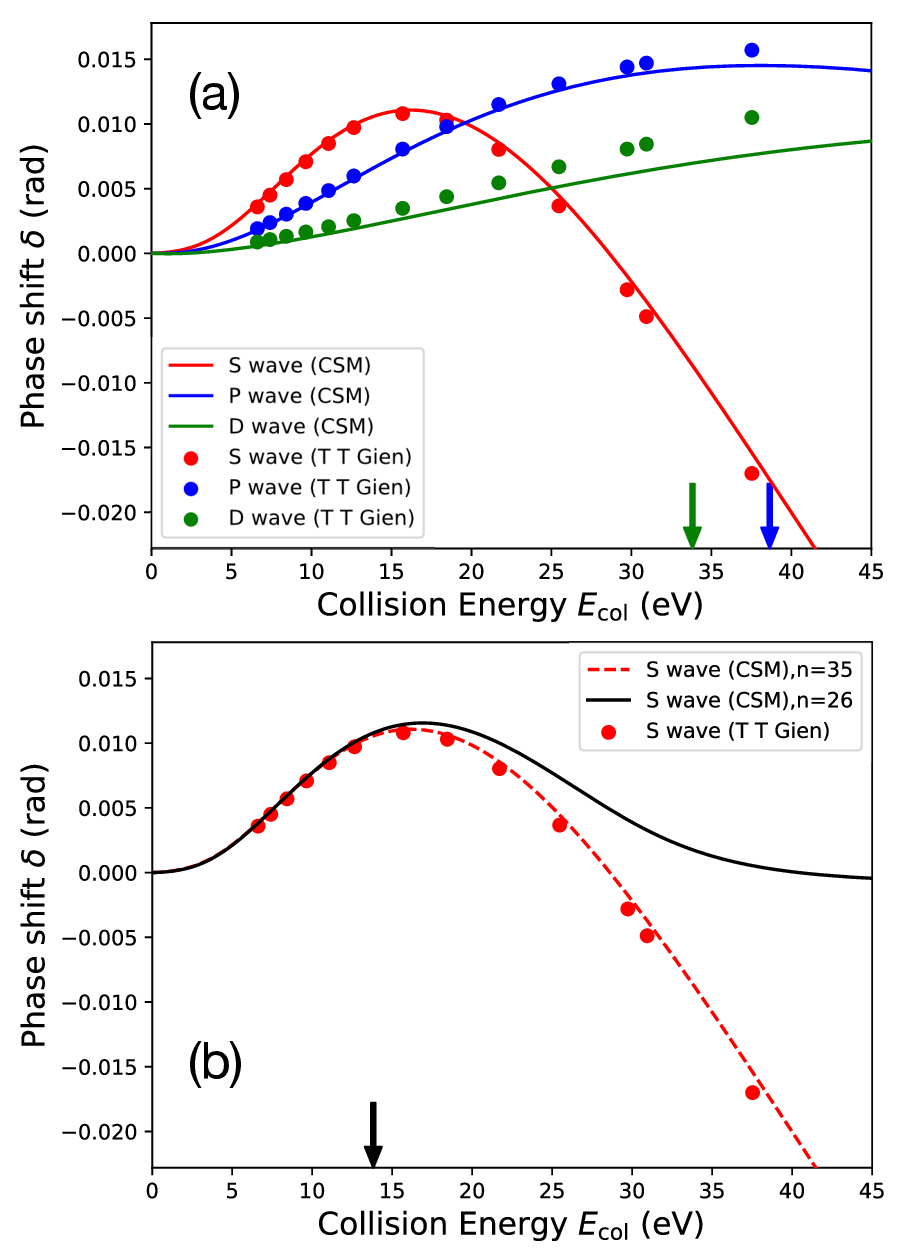}}
    \caption{(a) Calibrated phase shifts of the $\mathrm{e}^++\mathrm{He}^+$ scattering compared 
with those calculated using the Harris--Nesbet variational method~\cite{TTGien_2001}. 
For each partial wave, we used 35, 28, and 27 pairs of complex eigenenergies for the S-, P-, and D-waves, respectively. 
The arrow of the color corresponding to the partial wave presents $\mathrm{Re}E_{0,n}/2-E_\mathrm{He^+(1s)}$ as an indicator of the maximum applicable energy. 
The maximum $\mathrm{Re}E_{0,n}/2-E_\mathrm{He^+(1s)}$ for the S-wave is located at a much larger energy (approximately 116 eV). 
(b) The calibrated S-wave phase shifts of the $\mathrm{e}^++\mathrm{He}^+$ scattering calculated using the 35 pairs of complex eigenenergies are compared with 
those calculated using 26 pairs of complex eigenenergies. 
The black arrow presents $\mathrm{Re}E_{0,k=26}/2-E_\mathrm{He^+(1s)}$, which indicates the expected approximation limits for the collision energy.} 
    \label{fig.three-body-phase-shift}
\end{figure}

Figure~\ref{fig.three-body-phase-shift}(a) presents a comparison of the S-, P-, and D-wave phase shifts of e$^+$+He$^+(1s)$ 
with the latest calculations obtained using the Harris--Nesbet variational method~\cite{TTGien_2001}. We included a limited number of eigenenergy pairs (35, 28, and 27 pairs for S-, P-, and D-waves, respectively) and applied the calibration modification for low-energy scattering using Eq.~(\ref{eq.delta_k_dif_approx}).
The offsets caused by the modification are $-0.01953$ rad (S-wave), $0.00095$ rad (P-wave), and $0.00310$ rad (D-wave).
In this calculation, the indicators $\mathrm{Re}\,E_{0,n}/2-E_\mathrm{He^+(1s)}$ become 116 eV (S-wave), 38 eV (P-wave), and 32 eV (D-wave). 
It can be observed that the phase shifts calculated using the complex eigenenergies agree well with the Harris--Nesbet variational calculations.
While the P- and D-wave phase shifts present relatively large deviations from the reference at high energies in comparison with the S-wave phase shifts, 
which is consistent with the smaller $\mathrm{Re}\,E_{0,n}/2-E_\mathrm{He^+(1s)}$ than that of the S-wave.

Figure~\ref{fig.three-body-phase-shift}(b) presents the S-wave phase shifts of e$^+$-He$^+$ scattering to investigate the number of eigenenergy pairs included in the calculation. When the number of pairs is reduced from 35 to 26, the low-energy behavior of the phase shift remains unchanged, and the high-energy behavior does not match the precise calculation. The 26th eigenenergy corresponds to $\mathrm{Re}\,E_{0,n}/2-E_\mathrm{He^+(1s)}=14$ eV, which is approximately the maximum energy below which the calculated phase shifts are reliable.
This indicator appears to work well for the proposed three-body systems. The advantage of the calibration modification is that only a limited number of eigenenergies are required to determine the phase-shift behavior in low-energy scattering.

\begin{figure}[t]
    \centering
    \resizebox{0.48\textwidth}{!}{\includegraphics[angle=0]{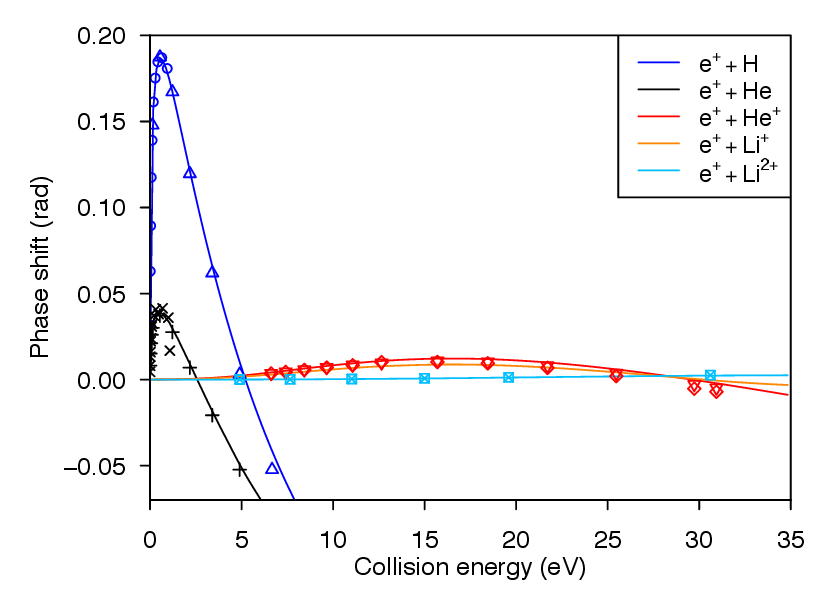}}
    \caption{S-wave phase shifts of positron scattering off H, He, He$^+$, Li$^+$, and Li$^{2+}$ atoms/ions are calculated using the CSM with calibration modification (solid lines). Points denote the previous works: $\circ$~\cite{PhysRevA.78.012703} and $\triangle$~\cite{PhysRevA.3.1328} for e$^+$-H, $+$~\cite{PhysRevA.90.032712} and $\times$~\cite{PhysRevA.78.012703} for e$^+$-He, $\diamondsuit$~\cite{Bransden_2001} and $\bigtriangledown$~\cite{TTGien_2001} for e$^+$-He$^+$, and $\boxtimes$~\cite{TTGien_2001b} for e$^+$-Li$^{2+}$.}
    \label{compare_various}
\end{figure}

Figure~\ref{compare_various} summarizes the S-wave phase shifts of e$^+$-X (X=H, He, He$^+$, Li$^+$, and Li$^{2+}$) scattering calculated using the CSM with a calibration modification. 
In general, the phase shift increases as the collision energy increases from zero and then decreases. The low-energy increase mainly originates from the attractive induced dipole interaction between the target atom/ion and the positron, and the decrease at higher energies is attributed to the positron-nucleus repulsive interaction. 
As observed in this figure, the phase shift behavior of the positron-neutral atom scattering significantly differs from that of the positron-ion scattering owing to the presence of a long-range repulsive Coulomb force in the latter case. 
We emphasize that the CSM calculation successfully describes these different phase-shift behavior and reproduces the results of previous studies using different methods. 

We also predicted the phase shift of e$^+$-Li$^+$ scattering using the same quality of calculation as e$^+$-He scattering. 
The obtained phase shift exhibited a similar but more moderate behavior than e$^+$-He$^+$, 
which is convincing because an increase in the nuclear charge prevents the positrons from approaching the electrons and reduces the attractive interactions between them. A similar trend can be observed in e$^+$-H and e$^+$-He, where the latter case shows more moderate behavior.

\begin{table*}[t]
\begin{ruledtabular}                
\caption{\label{tab:He+Li+}S-wave phase shifts of e$^+$+He$^+$ and e$^+$+Li$^+$ collisions. $x[y]$ denotes $x\times10^y$. Literature values are compared for e$^+$+He$^+$. 
Our calculations are noted as CSM$_{l_\mathrm{max}}$, where $l_\mathrm{max}$ is the maximum inner angular momentum used for the basis functions in this work.}
\begin{tabular}{cccccccccc}
    & \multicolumn{8}{c}{e$^+$+He$^+$} & \multicolumn{1}{c}{e$^+$+Li$^+$} \\ \cline{2-9}\cline{10-10}
$E_\mathrm{col}$(eV)  &  CSM$_{3}$  & CSM$_{2}$  &  CSM$_{1}$  &  CSM$_{0}$  &  Gien~\cite{TTGien_2001}   & Bransden~\cite{Bransden_2001}  &  Khan~\cite{Khan1984}  & Shimamura~\cite{doi:10.1143/JPSJ.31.217}  &  CSM \\ \hline
$6.62$   &   $4.56[-3]$  &   $4.56[-3]$  &   $4.48[-3]$  &   $1.42[-3]$    &   $3.59[-3]$    &  $3.56[-3]$  &    $3.40[-3]$  &   $1.50[-3]$  &   $3.23[-3]$ \\
$7.41$   &   $5.61[-3]$  &   $5.60[-3]$  &   $5.50[-3]$  &   $1.78[-3]$    &   $4.50[-3]$    &  $4.46[-3]$  &    $4.20[-3]$  &   $2.40[-3]$  &   $3.94[-3]$ \\
$8.43$   &   $6.97[-3]$  &   $6.97[-3]$  &   $6.85[-3]$  &   $2.27[-3]$    &   $5.70[-3]$    &  $5.60[-3]$  &    $5.20[-3]$  &   $4.00[-3]$  &   $4.92[-3]$ \\
$9.65$   &   $8.55[-3]$  &   $8.55[-3]$  &   $8.41[-3]$  &   $2.85[-3]$    &   $7.08[-3]$    &  $6.95[-3]$  &    $6.40[-3]$  &   $5.60[-3]$  &   $6.06[-3]$ \\
$11.07$  &   $1.02[-2]$  &   $1.02[-2]$  &   $1.00[-2]$  &   $3.45[-3]$    &   $8.49[-3]$    &  $8.29[-3]$  &    $7.40[-3]$  &   $7.00[-3]$  &   $7.23[-3]$ \\
$12.65$  &   $1.17[-2]$  &   $1.17[-2]$  &   $1.15[-2]$  &   $3.93[-3]$    &   $9.72[-3]$    &  $1.01[-2]$  &    $8.00[-3]$  &   $7.80[-3]$  &   $8.28[-3]$ \\
$15.7$   &   $1.33[-2]$  &   $1.33[-2]$  &   $1.31[-2]$  &   $4.10[-3]$    &   $1.08[-2]$    &  $1.02[-2]$  &    $7.90[-3]$  &   $6.60[-3]$  &   $9.35[-3]$ \\
$18.45$  &   $1.33[-2]$  &   $1.33[-2]$  &   $1.31[-2]$  &   $3.28[-3]$    &   $1.03[-2]$    &  $9.42[-3]$  &    $6.30[-3]$  &   $6.30[-3]$  &   $9.18[-3]$ \\
$21.71$  &   $1.17[-2]$  &   $1.16[-2]$  &   $1.15[-2]$  &   $1.16[-3]$    &   $8.03[-3]$    &  $6.80[-3]$  &    $2.70[-3]$  &   $4.40[-3]$  &   $7.67[-3]$ \\
$25.47$  &   $8.06[-3]$  &   $8.05[-3]$  &   $7.86[-3]$  &  $-2.61[-3]$    &   $3.67[-3]$    &  $2.01[-3]$  &    --          &   $6.00[-4]$  &   $4.55[-3]$ \\
$29.73$  &   $2.44[-3]$  &   $2.43[-3]$  &   $2.24[-3]$  &  $-8.13[-3]$    &  $-2.81[-3]$    &  $-5.32[-3]$  &  $-1.07[-2]$  &  $-5.40[-3]$  &  $-1.65[-4]$ \\
$30.94$  &   $6.42[-4]$  &   $6.32[-4]$  &   $4.37[-4]$  &  $-9.87[-3]$    &  $-4.88[-3]$    &  $-7.10[-3]$  &  $-1.31[-2]$  &  $-1.03[-2]$  &  $-1.65[-3]$ \\
$37.53$  &  $-1.00[-2]$  &  $-1.00[-2]$  &  $-1.03[-2]$  &  $-2.02[-2]$    &  $-1.70[-2]$    &  $-1.97[-2]$  &  $-2.68[-2]$  &  $-2.18[-2]$  &  $-9.43[-3]$
\end{tabular}
\end{ruledtabular}
\end{table*}

\begin{table}[t]
\begin{ruledtabular}                
\caption{\label{tab:Li2+}S-wave phase shifts of e$^+$+Li$^{2+}$ collision is compared with the literature values. Our calculations are noted as CSM$_{l_\mathrm{max}}$. 
CIKOHN$_8$ is referred to for Ref.~\cite{Novikov2004}. }
\begin{tabular}{cccccc}
$k$ (a.u.)    &  CSM$_3$ & CSM$_2$ & CSM$_1$ & Novikov~\cite{Novikov2004} &  Gien~\cite{TTGien_2001b} \\ \hline
$0.5$   &   $1.35[-4]$   &   $1.36[-4]$   &   $1.28[-4]$   &   $1.48[-5]$   &                \\    
$0.6$   &   $2.17[-4]$   &   $2.17[-4]$   &   $2.06[-4]$   &                & $3.68[-5]$     \\ 
$0.75$  &   $4.12[-4]$   &   $4.12[-4]$   &   $3.93[-4]$   &                & $1.20[-4]$     \\
$0.9$   &   $7.49[-4]$   &   $7.48[-4]$   &   $7.17[-4]$   &                & $3.13[-4]$     \\
$1.0$   &   $1.09[-3]$   &   $1.09[-3]$   &   $1.04[-3]$   &   $5.39[-4]$   &                \\
$1.05$  &   $1.30[-3]$   &   $1.29[-3]$   &   $1.25[-3]$   &                & $6.79[-4]$     \\
$1.2$   &   $2.08[-3]$   &   $2.08[-3]$   &   $2.01[-3]$   &                & $1.24[-3]$     \\
$1.5$   &   $4.01[-3]$   &   $4.01[-3]$   &   $3.90[-3]$   &   $2.57[-3]$   & $2.55[-3]$   
\end{tabular}
\end{ruledtabular}
\end{table}

Tables~\ref{tab:He+Li+} and \ref{tab:Li2+} list the numerical values of the S-wave phase shifts of the positron-scattering off the positive ions.
Our calculation is denoted as CSM$_{l_\mathrm{max}}$, where $l_\mathrm{max}$ indicates the maximum inner angular momenta, $l$ and $L$ in Eq. ~(\ref{eq:3bdWF}) and (\ref{eq:4bdWF}), used for the basis functions. 
With an increase in $l_\mathrm{max}$, the phase shifts converge in the presented energy range for the e$^+$+He$^+$ and e$^+$+Li$^{2+}$ collisions.
In the case of e$^+$+He$^+$ collisions, the converged values are slightly larger than those reported in the latest literature ~\cite{TTGien_2001,Bransden_2001}.
Similar trends are observed in the case of the e$^+$+Li$^{2+}$ collisions. 
Our calculation of CSM$_3$ for e$^+$+Li$^{2+}$ resulted in phase shifts a few times larger than those reported in the literature ~\cite{Novikov2004,TTGien_2001b}.
The larger evaluated phase shift at low energies than that in the previous studies indicates that the interaction between the positron and ions is 
described in a more attractive manner in our calculations.
Because the size of our basis functions is relatively large (10,852 at maximum), the description could be improved 
by including virtual target excitation and virtual Ps formation in comparison with previous calculations.
Another possible reason for the deviation from the literature values is the insufficiency of the completeness relationship Eq.~(\ref{eq.completeness_approx}).
In two-body potential scattering, the positron scattering wavefunction, in principle, satisfies the completeness relationship.
However, in few-body scattering, the positron wavefunction is coupled with the target ion states, including their virtual excitation. 
The contributions of the virtually excited states of the target ion
depend on the collision energy, which can lead to an insufficiency in the completeness relationship.
Thus, more careful investigations are required to determine the origin of the small differences in the phase shifts between the CSM calculations and other methods.

\section{Conclusion}
\label{conclusion}

We presented the phase-shift calculation of e$^+$-X scattering using CSM for the targets X=Ne, Ar, Kr, Xe, H, He, He$^+$, Li$^{2+}$, and Li$^{+}$. 
We observed the empirical fact that a complex eigenenergy pair whose real part is much greater than the collision energy is 
required to achieve the convergence of the phase shift, even for low-energy scattering.
Although the necessity for these high-lying complex eigenenergies may result in difficulties in few-body problems,
the contribution of these high-lying complex eigenenergies to the phase shift can be regarded as a constant,
and we proposed a modification to the CSM calculation. 
The use of the modification is suggested by the simple geometrical considerations described in Subsection ~\ref{sec:modification}.
The validity of the modification is demonstrated by the positron scattering off X=Ne, Ar, Kr, Xe, H, He, He$^+$, and Li$^{2+}$, which is
in good agreement with the literature.
We predicted the phase shifts of the e$^+$-Li$^+$ four-body scattering problem.

It should be noted that the present formulation of the phase-shift calculation was employed only on the complex energy plane.
It is intriguing that the use of CSM converts the quantum mechanical scattering problem 
into a problem of simple geometry on a complex energy plane. 
This may provide new insights into scattering problems in future studies.
While the demonstrated cases presented in this study are limited to the elastic scattering of the system
having no resonance/bound state,
more general formulations for a variety of systems, including resonance scattering and/or inelastic scattering, will be focused on in future works.
Positronium scattering off atoms, ions, and molecules below its breakup threshold
that has been investigated in recent experiments~\cite{PhysRevLett.115.223201} and theoretical development~\cite{PhysRevA.78.012703,PhysRevA.101.042705,PhysRevA.103.022817,PhysRevA.107.042802} would also be 
in the scope of this approach.


\begin{acknowledgments}
The authors would like to thank Dr. Takayuki Myo (Osaka Institute of Technology, Japan) for their useful suggestions regarding the formulation. 
T.Y. is thankful for the 
financial support received from the JSPS KAKENHI Grant Number JP22K13980.
Y.K. is thankful for the
JSPS KAKENHI Grant Number 18H05461.
T.S. and T.Y. contributed equally to this study.

This work was partly achieved through the use of the supercomputer system at Hokkaido University, and at Kyushu University.
\end{acknowledgments}


\appendix

\section{Expansion of $\delta_k$ and explicit form of $a_k$ and $b_k$}

Here, we describe the explicit expressions for $a_k$ and $b_k$ in Eq. ~(\ref{eq.delta_k_dif_ak_bk}). 
$\delta_k$ in Eq.~(\ref{eq.delta_k}) can be written as 
\begin{equation}
    \delta_k=\mathrm{tan}^{-1} \left( \tan (2\theta) \frac{1+\Delta_{y,k}}{1+\tilde{\Delta}_{x,k}} \right).
\end{equation}
The CSM calculation using $\theta<\pi/4$ (typically approximately $\pi/12$) ensures that $\tan (2\theta)<1$. 
As explained in the text, $\Delta_{y,k}\ll 1$ and $\tilde{\Delta}_{x,k}\ll 1$.
The Taylor expansion is applied to $({1+\tilde{\Delta}_{x,k}})^{-1}=1-\tilde{\Delta}_{x,k}+\tilde{\Delta}_{x,k}^2-\cdots$
and the Maclaurin expansion of $\mathrm{tan}^{-1}(z(1+\Delta))$, where $z(1+\Delta)<1$ and $\Delta\ll 1$, 
\begin{align}
\mathrm{tan}^{-1}(z(1+\Delta))= \mathrm{tan}^{-1}(z)+\frac{z}{z^2+1}\Delta-\frac{z^3}{(z^2+1)^2}\Delta^2 + \cdots,
\label{app.arctan_expansion}
\end{align}
we have Eq.~(\ref{eq.delta_expansion}).
Thus, the explicit forms of $a_k$ and $b_k$ are given as 
\begin{align}
a_k& 
= \frac{1}{2}\mathrm{sin}(4\theta)\biggl\{-\Delta_{x,k}+\Delta_{0x,k}+\Delta_{y,k}-\Delta_{0y,k} \nonumber\\
&+\mathrm{cos}^2(2\theta)\left(\Delta^2_{x,k}-\Delta^{\prime 2}_{x,k}\right)
-\mathrm{cos}(4\theta)\left(\Delta_{x,k}\Delta_{y,k}-\Delta_{0x,k}\Delta_{0y,k}\right)\nonumber \\
&-\mathrm{sin}^2(2\theta)\left(\Delta^2_{y,k}-\Delta^{\prime 2}_{y,k}\right)\biggr\},
\end{align}
and
\begin{align}
b_k 
&= \frac{1}{2}\mathrm{sin}(4\theta)\biggl\{-2\mathrm{cos}^2(2\theta)\left(\Delta_{x,k}-\Delta_{0x,k}\right) \nonumber\\
&+\mathrm{cos}(4\theta)\left(\Delta_{y,k}-\Delta_{0y,k}\right)\biggr\}.
\end{align}

In the evaluation of $| a_k \mathrm{Re}\,E_{0,k} / b_k |$ of Eq.~(\ref{eq.ak_bk_estim}), 
there exists an exceptional case wherein $\Delta_{x,k}-\Delta_{0x,k}\approx\Delta_{y,k}-\Delta_{0y,k}$.
In this case, $b_k$ should be smaller than $a_i (i<k)$, 
and $\delta_{{\rm diff},k}$ would have negligible contribution to the sum of $a_i (i<k)$.


%

\end{document}